\documentclass[aps,prl,twocolumn,nopacs,superscriptaddress,amsmath,amssymb,groupedaddress]{revtex4}  
\usepackage{graphicx}  
\usepackage{dcolumn}   
\usepackage{bm}        
\usepackage{amssymb}   
\usepackage{times}
\usepackage{ulem}
\usepackage{color}
\makeatletter
\@ifundefined{textcolor}{}
{%
 \definecolor{BLACK}{gray}{0}
  \definecolor{GRAY}{gray}{0.6}
 \definecolor{WHITE}{gray}{1}
 \definecolor{RED}{rgb}{1,0,0}
 \definecolor{GREEN}{rgb}{0,1,0}
 \definecolor{BLUE}{rgb}{0,0,1}
 \definecolor{CYAN}{cmyk}{1,0,0,0}
 \definecolor{MAGENTA}{cmyk}{0,1,0,0}
 \definecolor{YELLOW}{cmyk}{0,0,1,0}
 \definecolor{ORANGE}{rgb}{0.9,0.5,0}
}

\newcommand{\rmd}{{\rm d}}
\newcommand{\rme}{{\rm e}}
\newcommand{\rmi}{{\rm i}}

\makeatother

\begin{document}
\let\vaccent=\v{
\global\long\def\gv#1{\ensuremath{\mbox{\boldmath\ensuremath{#1}}}}
\global\long\def\uv#1{\ensuremath{\mathbf{\hat{#1}}}}
\global\long\def\abs#1{\left| #1 \right|}
\global\long\def\avg#1{\left< #1 \right>}
\let\underdot=\d{
\global\long\def\dd#1#2{\frac{d^{2}#1}{d#2^{2}}}
\global\long\def\pd#1#2{\frac{\partial#1}{\partial#2}}
\global\long\def\pdd#1#2{\frac{\partial^{2}#1}{\partial#2^{2}}}
\global\long\def\pdc#1#2#3{\left( \frac{\partial#1}{\partial#2}\right)_{#3}}
\global\long\def\op#1{\hat{\mathrm{#1}}}
\global\long\def\ket#1{\left| #1 \right>}
\global\long\def\bra#1{\left< #1 \right|}
\global\long\def\braket#1#2{\left< #1 \vphantom{#2}\right| \left. #2 \vphantom{#1}\right>}
\global\long\def\matrixel#1#2#3{\left< #1 \vphantom{#2#3}\right| #2 \left| #3 \vphantom{#1#2}\right>}
\global\long\def\av#1{\left\langle #1 \right\rangle }
 \global\long\def\com#1#2{\left[#1,#2\right]}
\global\long\def\acom#1#2{\left\{  #1,#2\right\}  }
\global\long\def\grad#1{\gv{\nabla} #1}
\let\divsymb=\div 
\global\long\def\div#1{\gv{\nabla} \cdot#1}
\global\long\def\curl#1{\gv{\nabla} \times#1}
\let\baraccent=\={

\title{Stabilizing arrays of photonic cat states via spontaneous symmetry breaking}


\author{Jos\'{e} Lebreuilly}
\affiliation{Laboratoire  Pierre  Aigrain, \'Ecole Normale  Sup\'{e}rieure -  PSL Research  University, CNRS,  Universit\'{e}  Pierre  et  Marie  Curie-Sorbonne  Universit\'{e}, Universit\'{e}  Paris  Diderot-Sorbonne  Paris  Cit\'{e},  Paris 75005, France}
\author{Camille Aron}
\affiliation{Laboratoire de Physique Th\'eorique, \'Ecole Normale Sup\'erieure, CNRS, PSL University, Sorbonne Universit\'e, Paris 75005, France}
\affiliation{Instituut voor Theoretische Fysica, KU Leuven, Belgium}
\author{Christophe Mora}
\affiliation{Laboratoire  Pierre  Aigrain, \'Ecole Normale  Sup\'{e}rieure -  PSL Research  University, CNRS,  Universit\'{e}  Pierre  et  Marie  Curie-Sorbonne  Universit\'{e}, Universit\'{e}  Paris  Diderot-Sorbonne  Paris  Cit\'{e},  Paris 75005, France}


\begin{abstract}
The controlled generation and the protection of entanglement 
is key to quantum simulation and quantum computation.
At the single-mode level, protocols based on photonic cat states hold strong promise as they present unprecedentedly long-lived coherence and may be combined with powerful error correction schemes. Here, we demonstrate that robust ensembles of ``many-body photonic cat states'' can be generated in a Bose-Hubbard model with pair hopping via a spontaneous $U(1)$ symmetry breaking mechanism. We identify a parameter region where the ground state is a massively degenerate manifold consisting of local cat states which are factorized throughout the lattice and whose conserved individual parities can be used to make a register of qubits. This  phenomenology occurs for arbitrary system sizes or geometries, as soon as long-range order is established, and it extends to driven-dissipative conditions. In the thermodynamic limit, it is related to a Mott insulator to pair-superfluid phase transition.
\end{abstract}

\maketitle

\paragraph*{Introduction.}The ability to engineer a large variety of Hamiltonian couplings with sufficient tunability and to protect quantum coherence is essential for the generation of exotic quantum states~\cite{Bloch_rev,Houck_rev,Carusotto_topo} and for quantum information processing~\cite{Haroche_rev,Devoret_rev}.
In various platforms such as ultracold atomic gases or superconducting circuits, the huge timescale separation between Hamiltonian and dissipative dynamics is suitable for adiabatic preparation schemes where the physics is dominated by ground-state properties. Adopting a new perspective, quantum reservoir engineering ideas~\cite{Zoller_QRE,Verstraete_QRE,Diehl_atom_dissipation,Rydberg_reservoir_engineering,Optomechanical_reservoir_engineering} that harness  dissipation as a resource rather than a flaw have opened the gates to mostly uncharted territory: the nonequilibrium generation of quantum states.

In the context of cavity-QED, quantum reservoir engineering schemes have been proposed and successfully implemented for the preparation of single qubit states on the Bloch sphere~\cite{Murch2012}, entangled states of distant qubits~\cite{Shankar2013, Lin2013, Mollie2016,Liu2016}, and very recently of the first Mott insulator of light~\cite{Mott_exp,Lebreuilly_2016,Ma_Simon,Alberto_BH,Lebreuilly_square}. Another resounding success for quantum error correction
 is the preparation of photonic cat states (PCS)~\cite{Encoding_cat,Leghtas_cat,Girvin_cat_code},
\begin{equation}
\ket{\mathcal{C}^{\pm}(\alpha)}\propto\left(\ket{\alpha}\pm\ket{-\alpha}\right)\,,
\end{equation}
which are macroscopic multi-photon superpositions naturally insensitive to dephasing in the limit of a large $\alpha$, and which can be efficiently protected against photon losses via parity measurement~\cite{Girvin_parity} and feedback control~\cite{Mirrahimi_cat,Bartolo_cat,Blais_Cat}.
These schemes heavily rely upon the current development of generic nonlinearities of the type $a^{n \dagger} b^m$ with $n,b  \in \mathbb{N}^*$, such as 3-wave mixing \textit{i.e.} $n+m=3$~\cite{Raman6_Devoret,SNAIL, SNAIL_2}.

A natural extension of this joint endeavour is the preparation of cat states in the many-body context. However, the conditions for the emergence of cat states in large multimode architectures still remains elusive, as hybridization between neighboring PCS is expected to be detrimental to the local nature of those states. In this Letter, we claim that large quantum registers of PCS can be spontaneously generated by bringing extended bosonic systems to develop a pair-superfluid (PSF) order~\cite{PSF_Ripoll,PSF_bonnes}, and we propose both equilibrium and driven-dissipative routes for the emergence of this many-body phase.
The intimate connection between PCS and a PSF phase (the bosonic counterpart of the Barden-Cooper-Schrieffer, BCS, phase), starts with them sharing identical symmetries: they both break $U(1)$ invariance while preserving a $\mathbb{Z}_2$ subsymmetry. Until now, generating individual  PCS has mostly been achieved by \emph{explicitly} breaking the $U(1)$ symmetry associated with particle number conservation, \textit{e.g.}, by shining single- or two-photon coherent sources on photonic cavities~\cite{Leghtas_cat,Bartolo_cat,Blais_Cat,Clerk_cat,Savona_SSB_Z_new}. Here, we shall rather capitalize on  a purely many-body mechanism, namely a \emph{spontaneous symmetry breaking} of $U(1)$ invariance, to achieve the long-range PSF order and generate arrays of PCS.

\begin{figure*}[t]
\centering
\includegraphics[width=1\textwidth,clip]{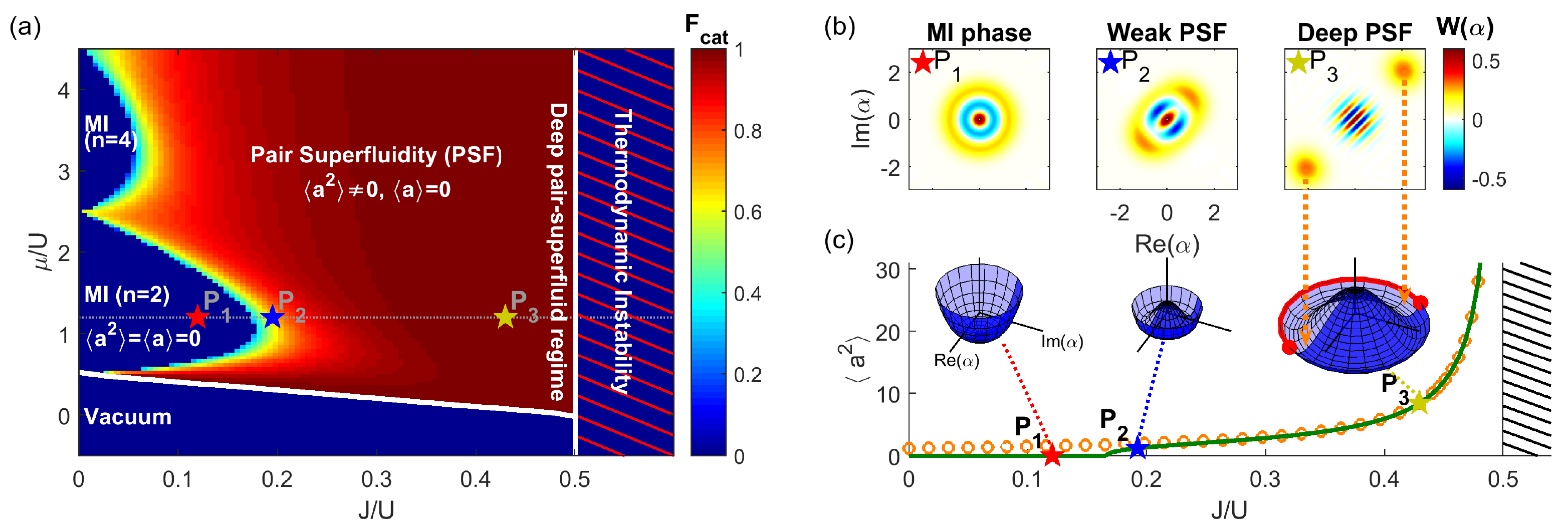} 
\hspace{1cm}
\caption{(a): Zero-temperature phase diagram in the even-parity sector as a function of the pair-hopping $J$ and the chemical potential $\mu$ in units of the interaction $U$. The fidelity $F_{\rm{cat}}=\max_{\alpha}(|\langle \psi_{\rm{GW}}|\mathcal{C}^{+(\alpha)}\rangle|^2)$ between the Gutzwiller wave function and the closest cat state is represented in color plot. (b): Wigner quasiprobability distribution~\cite{Haroche} of the Gutzwiller wave function $\ket{\psi_{\rm{GW}}}$ for three points in the phase diagram, respectively located in the Mott Phase $(P_1)$, and in the pair-superfluid phase in the weak  $(P_2)$ and strong $(P_3)$ PSF regimes. (c): pair-superfluid order parameter as a function of $J$ at fixed chemical potential $\mu/U=1.2$. The green solid line and the orange circles represent the predictions of the Gutzwiller analysis and the semiclassical result of Eq.~(\ref{eq:ground-state_density}), respectively. 
In the inset, the formation of a Mexican hat potential sketches the spontaneous symmetry-breaking mechanism generating the many-body photonic cat states. 
\label{fig:Gutzwiller_equilibrium}}
\end{figure*}
\paragraph*{Equilibrium model.} We flesh out our proposal in the context of a modified Bose-Hubbard model
\begin{equation}
H_{0}=\frac{U}{2}\sum_i a_{i}^{\dagger 2} a_{i}^2-\frac{J}{z}\sum_{\langle {i},{j}\rangle} \left[a_{i}^{\dagger 2} a_{j}^2 + \mathrm{H.c.} \right]\,,\label{eq:extended_BH}
\end{equation}
where the usual single-particle hopping between a site $i$ and its $z$ nearest neighbors is replaced by two-particle hopping processes of amplitude $J$, and $U>0$ accounts for standard on-site repulsive interactions. Such pair interactions can be readily realized within ultracold atoms~\cite{PSF_Ripoll}  or circuit-QED~\cite{Leghtas_cat} platforms. We propose and detail a realistic implementation for the latter in the Supplemental Material~\cite{SM}.
The effects of lattice geometry, such as the distinction between a superfluid order and Bose-Einstein condensation in low dimensions, is washed away in both our subsequent mean-field description and exact results. 
 However, for simplicity purposes we have in mind a cubic lattice with $N_{\rm{sites}}$ and periodic boundary conditions. In addition to the \emph{global} $U(1)$ symmetry, $a_i \mapsto \rme^{\rmi \theta} a_i$, corresponding to the conservation of the total number of particles $N$, the Hamiltonian is also symmetric under \emph{local} discrete $\mathbb{Z}_2^{\rm loc}$ transformations, namely $a_i \mapsto \zeta(i) a_i$ with $\zeta(i) = \pm$ where $\zeta(i)$ can vary from site to site. This latter symmetry corresponds to the conservation of the parity of the particle number at each site. 
\paragraph*{Ground-state phase diagram.} The zero-temperature phase diagram was obtained numerically within a Gutzwiller mean-field approach (see Suppl. Mat.~\cite{SM}). We monitored both the single-particle and two-particle order parameters, $\psi^{(1)} \equiv \langle a_i \rangle$ and $\psi^{(2)} \equiv \langle a_i^2 \rangle$. Another important figure of merit is the fidelity $F_{\rm{cat}}\equiv\max_{\alpha,\pm}|\langle \psi_{\rm{GW}}|\mathcal{C}^{\pm}(\alpha)\rangle|^2$ between the local ground-state wavefunction $\ket{\psi_{\rm{GW}}}$ of the Gutzwiller ansatz and the closest cat state.
The outcome is displayed in Fig.~\ref{fig:Gutzwiller_equilibrium}a as a function of the chemical potential $\mu$ and the pair-hopping $J$. For simplicity, we restricted the results to the sector with only even local parities, see Suppl. Mat.~\cite{SM} for a complete picture including the odd parity sector.

At weak hopping amplitude $J$, the ground state is analogous to the one of the standard single-particle hopping Bose-Hubbard model. It features a series of Mott-insulating regions with even integer densities $n\equiv \langle a_{i}^\dagger a_i \rangle=0, 2,4, \ldots$ characterized by $\psi^{(1)} =  \psi^{(2)} = 0$, reflecting the underlying global $U(1)$ symmetry. The lobe boundaries at stronger $J$ correspond to a second-order phase transition to a superfluid phase where the $U(1)$ symmetry is spontaneously broken while the $\mathbb{Z}_2$ symmetry is preserved. Here, given that superfluidity is only carried by pairs of photons, this translates into a vanishing single-particle order parameter $\psi^{(1)} = 0$ and a non-vanishing two-particle order parameter $\psi^{(2)} \neq 0$.

As $J$ is increased towards the special value $J_\ast\equiv U/2$, the PSF order parameter $\psi^{(2)}$ diverges to $+\infty$, and the local fidelity to a cat state approaches one. The corresponding Wigner functions, displayed in Fig.~\ref{fig:Gutzwiller_equilibrium}, illustrate the continuous change via spontaneous symmetry breaking from a Fock state in the Mott-insulating phase to a cat state deep into the PSF phase. Finally, we emphasize that the protection of the global $\mathbb{Z}_2$ symmetry can survive even in absence of the local $\mathbb{Z}_2^{\rm loc}$ symmetries, as a consequence of a non-zero energy gap $\Delta\propto\exp{(-C/(U-2J))}$ separating the even and odd parity states for $J<U/2$ (see Suppl. Mat.~\cite{SM} where we illustrate in particular the robustness of PSF order and many-body PCS against single-particle hopping).
\paragraph*{Many-body cat states.}Remarkably, at the special value $J_\ast=U/2$, the gap $\Delta$ cancels exactly and one can analytically compute the ground state of the many-body Hamiltonian $H_0-\mu N$. At $\mu = 0$, the ground-state manifold is located at zero energy and spanned by the following set of many-body wave-functions,
\begin{equation}
\label{eq:ground-state_zero_mu}
\ket{\psi^{P}(\alpha)}=\bigotimes_{i} \ket{\mathcal{C}^{P({i})}(\alpha)}_{i}\,,
\end{equation}
defined as an extended product state of local PCS. The proof of this exact result is detailed in~\cite{SM}. The hopping-induced locking of the cat states at a common  coherent field $\alpha$ is the consequence of a protection against relative dephasing between the various lattice sites. Importantly, the on-site parity $P({i})=\pm$ can vary from site to site, yielding an extensively large degeneracy of the ground-state manifold. This latter property is particularly compelling for quantum memory applications, as the parities at each sites act as an emergent stable  register of qubits. Large-scale and versatile entanglement between these qubits can then be prepared via arbitrary superpositions of these states, which are themselves preserved by the many-body dynamics generated by $H_0$.

At finite chemical potential $\mu$ and $J=U/2$, the states of Eq.~(\ref{eq:ground-state_zero_mu}) are no longer eigenstates but rather follow the simple dynamical evolution $\ket{\psi(t)}=\ket{\psi^{P}(\alpha \rme^{-\rmi \mu t})}$. Moreover, defining $\mathcal{P}_N$ as the projector onto the submanifold with a total particle $N$, the exact ground states within this subspace are
 \begin{equation}
\label{eq:ground-state_finite_mu}
\ket{\psi^{P}_N}=\mathcal{P}_N\ket{\psi^{P}(\alpha)}\,
 \end{equation}
and are located at an energy $-\mu N$. The degeneracy associated to the local parities is preserved.
 The distinction between $\psi^{P}_N$ and $\psi^{P}(\alpha)$ is nonetheless meaningless in the thermodynamic limit ($N_{\rm sites}\to +\infty$). There, $\psi^{P}(\alpha)$ thus accurately describes the ground-state physical properties even for  $\mu \ne 0$.
Importantly for realistic implementations, these exact results are valid regardless of the system size and spatial dimensionality, and PCS are expected already with $N_{\rm sites} = 2$ sites (see also Ref.~\cite{Fischer_cat} where a similar phenomenology was observed for a two-mode system). A detection scheme of the $N$-particle many-body cat states $\ket{\psi^{P}_N}$ is detailed in the Supplemental Material~\cite{SM}.
\paragraph*{Thermodynamic instability and semiclassical analysis.}We emphasize that the exact solutions are located on the verge of an instability. This can be seen by using the coherent state $\bigotimes_i\ket{\alpha}$ as a variational ansatz, yielding an energy landscape $-\mu |\alpha|^2 +\left(U/2-J\right)|\alpha|^4$ which is unbounded from below for $J>U/2$.
On the contrary, for $J<U/2$ the model is thermodynamically stable as confirmed by our Gutzwiller numerical calculations and exact results (see Suppl. Mat.~\cite{SM} for the proof). In this case, a
first-order calculation of the ground-state energy in $J-U/2$ provides a precise estimate of the density and PSF order parameter close to the instability threshold
\begin{equation}
\label{eq:ground-state_density}
\langle a_{i}^\dagger a_{i}\rangle\simeq |\langle a_{i}^2\rangle|\underset{J\to U/2}{\simeq}\frac{\mu}{U-2J},
\end{equation}
regardless of the choice of on-site parities $P(i)$.
The excellent agreement of the semiclassical description with Gutzwiller simulations (see Fig.~\ref{fig:Gutzwiller_equilibrium}c)~\footnote{In contrast with the standard Bose-Hubbard model with single-particle hopping~\cite{Fisher}, the regime of validity of the semiclassical description is thus located at finite $J$ around $J_\ast=U/2$ rather than at infinite hopping.}, together with the strict local-parity conservation, further explain why the emergence of many-body cat states occurs throughout a wide region of the phase diagram, $U/3\lesssim J\leq U/2$ .

Beyond semiclassics, we compute the Bogoliubov spectrum of the elementary excitations preserving the local parities (the excitations corresponding to a change of local parity are characterized by the gap $\Delta$). We find the dispersion law $E_{\boldsymbol{k}}=\sqrt{\xi_{\boldsymbol{k}}(\xi_{\boldsymbol{k}}+2\mu)}$, typical of superfluidity, with a finite sound speed $c=2a\sqrt{2\mu J}$ at low-momenta and verifying the Landau criterion.  Here $\xi_{\boldsymbol{k}}\equiv -4J/z\sum_{\nu=1}^d[\text{cos}(k^\nu a)-1]$, and $a$ is the lattice constant. The validity of the  Landau criterion ensure that the physics is dominated by the ground-state manifold of many-body PCS even in presence of small perturbations to $H_0$.
\begin{figure}[t]
\centering
\includegraphics[width=0.9\columnwidth,clip]{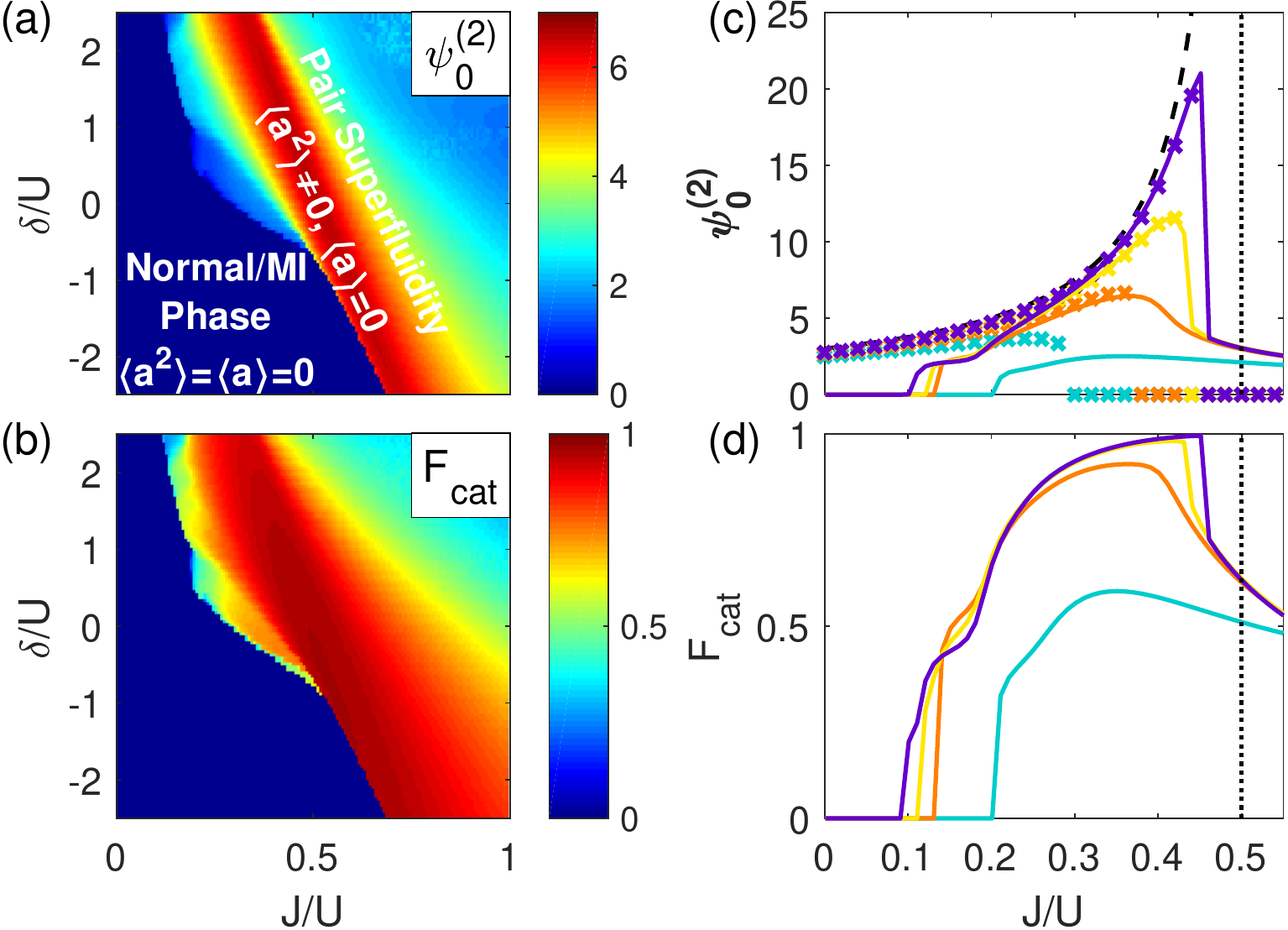} 
\hspace{1cm}
\caption{(a)-(b): Phase diagram in the driven-dissipative scenario, truncated to the even-parity sector, as a function of the pair-hopping $J$ and the detuning $\delta=\omega_{\rm at}/2-\omega_{\rm c}$ in units of the interaction $U$. The two-photon loss rate is set to $\Gamma_{\rm l}=10^{-2}\Gamma_{\rm p}$. The pair-superfluid order parameter $\psi_0^{(2)}=\langle a_i^2\rangle$, and the fidelity $F_{\rm{cat}}=\max_{\alpha}\langle  \mathcal{C}^{+(\alpha)}|\rho_{\rm{GW}}|\mathcal{C}^{+(\alpha)}\rangle$ between the Gutzwiller density matrix $\rho_{\rm GW}$ and the closest cat state are represented in color plot in panels (a) and (b), respectively. (c)-(d): $\psi_0^{(2)}$ and $F_{\rm{cat}}$ are respectively represented for various dissipative rates $\Gamma_{\rm{em}}^0,\Gamma_{\rm{l}}$ as a function of $J$ and at fixed detuning $\delta=2U$. The results of the Gutzwiller analysis are displayed in solid lines, and compared for $\psi_0^{(2)}$ to the semiclassical results with (resp. without) the effect of saturation, displayed in crosses (resp. black dashed line). All panels feature an interaction strength $U/\Gamma_{\rm p}=0.7$, and the Rabi coupling  $\Omega_{R}$ is chosen  to maintain a fixed ratio $\Gamma_{\rm{em}}^0/\Gamma_{\rm{l}}=9$ between pumping and losses. The various curves of panels (c) and (d) correspond to $\Gamma_{\rm l}/\Gamma_{\rm p}=3\times 10^{-2}$ (blue), $1\times 10^{-2}$ (orange), $3\times 10^{-3}$ (yellow), and $1\times 10^{-3}$ (purple). 
\label{fig:Gutzwiller_dissipative}}
\end{figure}
\paragraph*{Driven-dissipative model.}
Many-body systems in a spontaneously broken phase are naturally sensitive to external perturbations breaking the symmetry: in the Supplementary Material~\cite{SM}, we show how to prepare a large ensemble of cat states by simply shining a two-photon coherent drive at a single dissipative site in the lattice. Such a hardware-efficient method could be of interest for the implementation and the initialization of quantum registers.

In the remainder of this work, we rather investigate the connection between many-body PCS and spontaneously-broken PSF phases within a driven-dissipative scenario which preserves the initial $U(1)$ symmetry of the model as well as the conservation of local parities: to the unitary physics of the two-particle hopping Bose-Hubbard model, we add two-photon decay channels and incoherently pumped two-level systems exchanging pairs of photons.
 %
 
The dynamics are described by the following master equation:
\begin{multline}
\label{eq:master_equation}
\partial_t \rho=- \rmi \left[H_{\rm ph} + H_\mathrm{at} +H_\mathrm{ph-at} ,\rho\right]\\+\sum_{i}\left\{\Gamma_{\rm{l}}\mathcal{D}[a_{i}^2](\rho)+\Gamma_{\rm{p}}\mathcal{D}[\sigma_{i}^+](\rho)\right\} ,
\end{multline}
with the photonic Hamiltonian 
$H_{\rm ph} = H_0 + \sum_i \omega_{\rm c} a_i^\dagger a_i$.
$H_0$ is two-photon hopping Bose-Hubbard Hamiltonian previously introduced in Eq.~(\ref{eq:extended_BH}) and $\omega_{\rm c}$ is the cavity frequency. $H_\mathrm{at} = \sum_i  \omega_{\rm{at}}\sigma_{i}^{+}\sigma_{i}^-$  is the Hamitonian of the two-level systems which can coherently emit or absorb pairs of photons at a Rabi frequency $\Omega_{\rm{R}}$ according to $H_\mathrm{ph-at} =\Omega_{\rm{R}}\sum_{i} [\sigma_{i}^{-}a_{i}^{\dagger 2}+\sigma_{i}^{+}a_{i}^2]$. Finally the Lindblad superoperators  in the second line of Eq.~(\ref{eq:master_equation}) account for two-photon losses and an incoherent pumping of the two-level emitters occuring at rates $\Gamma_{\rm{l}}$ and $\Gamma_{\rm{p}}$ respectively. We used the notation $\mathcal{D}[X](\rho)\equiv X\rho X^{\dagger }-1/2\left\lbrace X^{\dagger }X,\rho\right\rbrace$.

The main function of the pumped two-level systems is to implement a frequency-dependent incoherent pump injecting photons by pairs~\cite{Lebreuilly_2016,Lebreuilly_2018} at a frequency-dependent rate
 \begin{equation}
 \label{eq:pump_spectrum}
\mathcal{S}_{\rm{em}}(\omega)=\Gamma_{\rm{em}}^0\frac{(\Gamma_{\rm{p}}/2)^{2}}{(\omega-\omega_{\rm{at}})^2+(\Gamma_{\rm{p}}/2)^{2}}\,,
\end{equation} 
whose maximum is set by $\Gamma_{\rm{em}}^0=4\Omega_{\rm{R}}^2/\Gamma_{\rm{p}}$. Our scheme is most efficient in the non-saturating regime ($\Gamma_{\rm{em}}^0\max(n^2,1)\ll \Gamma_{\rm{p}}$, with $n$ the density) where once a two-level system has emitted, it is quickly and efficiently pumped back to its excited state, thus maintaining a nearly perfect population inversion, and in the weakly dissipative regime ($\Gamma_{\rm{em}}^0$,  $\Gamma_{\rm{l}}\ll U, J$) where the photonic dynamics are dominated on short timescales by the Hamiltonian part $H_{\rm{ph}}$. Single-photon losses are detrimental to our scheme: we assume that these processes occur at a rate $\gamma\ll\Gamma_{\rm{em}}^0,\,\Gamma_{\rm{l}}$, 
such that there is enough time for the relaxation within each parity sector to take place  before any single-photon loss event occurs, and we study the physics within this transient regime.
\paragraph*{Steady-state phase diagram.}
After a relaxation period, the driven-dissipative dynamics are expected to reach a nonequilibrium steady state. Neglecting single-particle losses (see discussion above), the steady states are non-unique and present a large multiplicity: the state reached after a long evolution depends on the local parities initially imprinted on the system. Here we restrict ourselves to the even-parity sector, and explore the resulting phase diagram by means of a nonequilibrium Gutzwiller mean-field approach (see Suppl. Mat.~\cite{SM}). The results are presented in Fig.~\ref{fig:Gutzwiller_dissipative} as a function of the two-photon hopping $J$ and the detuning  $\delta \equiv \omega_{\rm{at}}/2-\omega_{\rm{c}}$  which, as we will see, plays a role analogous to the equilibrium chemical potential.

Similarly to the zero-temperature equilibrium case, we find a normal phase with $\psi^{(1)}(t) \equiv    \langle a \rangle(t) = 0$ and $\psi^{(2)}(t) \equiv  \langle a^2 \rangle(t)= 0$ at weak hopping $J$. For this particular computation ($U=0.7\,\Gamma_{\rm{p}}$), the resulting phase is not insulating. The stabilization of a photonic Mott insulator in the strong photon blockade regime ($U\gg\Gamma_{\rm{p}}$) is discussed in Refs.~\cite{Lebreuilly_2016,Lebreuilly_2018}. At stronger hopping amplitudes $J$, we find the onset of PSF: the $U(1)$ symmetry is spontaneously broken, yielding a non-vanishing two-photon order parameter
$\psi^{(2)} \ne 0$, while the single-photon order parameter remains zero, $\psi^{(1)}=0$, as a consequence of the unbroken global $\mathbb{Z}_2$ symmetry. 

Similarly to the equilibrium case, the steady-state Gutzwiller density matrix $\rho_{\rm{GW}}$ is found to be very close to a PCS in a wide region of the PSF phase, with fidelities achieving values over $95\%$ for a ratio $U/\Gamma_{\rm l}=70$
(see Fig.~\ref{fig:Gutzwiller_dissipative}a-b). Moreover, a scaling analysis presented in Fig.~\ref{fig:Gutzwiller_dissipative}c-d indicates that fidelity even reaches unity in the ideal limit of vanishing rates $\Gamma_{\rm{m}}^0,\, \Gamma_{\rm{l}}$.  We nonetheless highlight two strong differences with zero temperature, which we show below to proceed from the saturation of emitters. First, the two-photon field $\psi^{(2)}$ does not diverge and  presents an upper bound. Second, the domain of optimal fidelity to PCS, and maximum $\psi_{\max}^{(2)}$, is tilted with respect to  $J/U = 1/2$.  

\paragraph*{Semiclassical analysis.}
In order to gain further insight on the complex driven-dissipative dynamics of our model, we derived the self-consistent mean-field equation on the order parameter $\psi^{(2)}(t) $ in the semiclassical regime $|\psi^{(2)}(t)|\gg1$ (see details in Suppl. Mat.~\cite{SM}).

In agreement with the Gutzwiller results, the PSF order parameter develops a steady-state oscillatory behavior in the $U(1)$-broken phase,  $\psi^{(2)}(t) = \psi_0^{(2)} \rme^{-\rm i \omega_{\mathrm{PSF}} t}$. Non-trivial solutions are found only above the lasing threshold $\Gamma_{\rm{em}}^0\geq \Gamma_{\rm l}$,  with their amplitude obeying 
\begin{equation}
\label{eq:SF-fraction_MF}
\psi_0^{(2)}=\frac{\delta_{\rm PSF}}{U-2J}.
\end{equation}
The effective detuning $\delta_{\rm PSF}\equiv \omega_{\rm PSF}/2-\omega_{\rm{c}}$ and the order parameter frequency $\omega_{\rm PSF}$ are set by the balance between the energy injected in the photonic system and the energy lost by dissipation
\begin{equation}
\label{eq:SF-frequency_sat_2}
\Gamma_{\rm{l}}=\frac{\mathcal{S}_{\rm{em}}(\omega_{\rm PSF})}{1+s}\,,
\end{equation}
where the saturation parameter $s=2|\psi_0^{(2)}|^2\mathcal{S}_{\rm{em}}(\omega_{\rm PSF})/\Gamma_{\rm{p}}$ limits the pump amplification power. The Lorentzian form of $\mathcal{S}_{\rm{em}}$ then yields two distinct frequencies $\omega_{\rm PSF}$, however at most one solution  at a time was found to be non-trivial, physical and dynamically stable. 

Noteworthy, observe the strong similarity between Eq.~\eqref{eq:SF-fraction_MF} and its equilibrium counterpart in Eq.~\eqref{eq:ground-state_density} when identifying the effective detuning $\delta_{\rm PSF}$ with the chemical potential $\mu$. The divergence in $\psi_0^{(2)}$ can be interpreted as a breakdown of the photon blockade at $J_\ast=U/2$: the energy separation $2\omega_{\rm{cav}}$ between two successive groundstates $\ket{\psi_N^P}$ and $\ket{\psi_{N+2}^P}$ of $H_{\rm{cav}}$ does not depend anymore on the total particle number $N$, as the two-photon hopping counterbalances perfectly the photon repulsion. However, as shown in Eq.~\eqref{eq:SF-frequency_sat_2}, when $\psi^{(2)}_{0}$ increases the saturation becomes relevant, setting an upper bound to the order parameter $\psi^{(2)}_{\max}=\sqrt{\left(\Gamma_{\rm{p}}/\Gamma_{\rm{l}}-\Gamma_{\rm{p}}/\Gamma_{\rm{em}}^0\right)/2}$. $\psi^{(2)}_{\max}$ is achieved at $J_{c}=U/2-\delta/(2\psi^{(2)}_{\max})$ which depends linearly on $\delta$. This explains the tilting of the PSF domain observed in  the Gutzwiller computations, as well as $J_{c}=U/2$ at $\delta=0$. 

The results of the semiclassical analysis are presented in Fig.~\ref{fig:Gutzwiller_dissipative}c (crosses) for various degrees of saturation, and accurately reproduce the Gutzwiller predictions (solid lines) within its regime of validity, \textit{i.e.}, when $|\psi^{(2)}_{0}| \gg 1$. As predicted, in the limit of a vanishing photon pumping rate $\Gamma_{\rm{em}}^0/\Gamma_{\rm p}\to 0$ at fixed ratio $\Gamma_{\rm{em}}^0/\Gamma_{\rm{l}}$, both in the Gutzwiller and semiclassical results (black dashed lines) predict a diverging order parameter $\psi^{(2)}_{\max}\to +\infty$ as well as an instability located at $J_{c}=U/2$, even for a non-vanishing detuning $\delta$.
\paragraph*{Conclusions.} 

In this work, we developed a comprehensive theoretical framework for the emergence of photonic cat states in the many-body context, and the
preparation of large ensembles of these states via spontaneous symmetry breaking. Questions left opened are the precise characterization of the elementary excitations in the driven-dissipative scenario. From another perspective, our work suggests a non-conventional path for continuous-variable quantum computing~\cite{Cont_var_QC} taking advantage of many-body effects to protect quantum coherence. The phenomenology presented here can be generalized to four-photon physics~\cite{Mirrahimi_cat,Girvin_cat_code} in view of correcting the dynamics against single-photon loss events.
\begin{acknowledgments}\textit{Acknowledgements.} The authors thank Iacopo Carusotto, Fabrizio Minganti,  Steven Girvin, Nicholas Frattini, and Alexander Grimm for stimulating discussions.
\end{acknowledgments}

\bibliographystyle{apsrev4-1}
\bibliography{Bibliography}
\onecolumngrid
\newpage
\begin{center}
{\Large Stabilizing arrays of photonic cat states via spontaneous symmetry breaking \\ Supplementary Material}
\end{center}
\section*{\uppercase{Circuit-QED implementation of pair hopping}}
In this Supplementary Material, we propose a realistic implementation of the Bose-Hubbard model with pair hopping in a superconducting-circuit architecture.
In this context, the realization of multi-photon couplings typically involves a combination of nonlinear inducting dipoles and parametric modulations.

Note that one could think of simply using Josephson junction couplings to realize coherent transport of pairs of photons while suppressing single-photon hopping: in the single-photon case, a beam splitter between two resonators of different frequencies $\omega_{1}$ and $\omega_{2}$ can be realized via a Josephson junction parametrically modulated at the frequency difference $\omega_{2}-\omega_{1}$. Instead, modulating at $2(\omega_{2}-\omega_{1})$ would enable only the two-photon hopping processes and leave the single-photon hopping strongly off-resonance. 
However, in addition to providing the desired coupling, the use of four-wave mixing Josephson elements usually yields spurious couplings, some of which often reveal detrimental. In our case, such an approach would lead in particular to the presence of negative self-Kerr and cross-Kerr terms~\cite{QL_ampli,Leghtas_cat,QEC_binomial}, which in the many-body language translate to on-site and nearest neighbor attractive interactions, respectively. Note also that while one can argue that the physics described in the manuscript could also be observed in presence of negative interactions (by evolving adiabatically in the highest-energy manifold in the isolated case instead of the ground-state manifold, or by simply changing the sign of the detuning between the cavities and the emitters in the driven-dissipative case), such a modification of the original model goes beyond the scope of our work. 

Here instead, we propose an alternative method for the generation of multi-photon transport processes circumventing the effects mentioned above. Beyond many-body cat states, the development of methods to avoid cross-Kerr couplings is of interest for quantum simulation applications. Our approach, described in Fig.~\ref{fig_supp:pair-hopping}, is based on the recently developed SNAIL devices~\cite{SNAIL,SNAIL_2} (Superconducting Nonlinear Inductive Asymmetric eLement) presenting important three-wave mixing, as well as tunable and possibly cancellable four-wave mixing. The lattice sites are inspired by the design of fluxonium qubits, although they operate in a different parameter range. The flux injected in the qubit allows to tune the strength and the sign of the two-photon interactions $U_0$, and to even completely cancel it if desired.
We assume $U_0>0$. The qubits are coupled virtually via three-wave mixing processes to auxiliary resonators, whose frequency $\omega_{\rm{aux}}=2\omega_{\rm{c}}+\Delta$ (with $\Delta\ll\omega_{\rm{c}}$) is close to twice the bare frequency $\omega_{\rm{c}}$ of the main lattice sites. The presence of this auxiliary degree of freedom efficiently prevents cross-Kerr couplings between the main lattice sites.

\medskip

Below, we 
\begin{itemize}
	\item present the second-quantized many-body Hamiltonian describing the circuit;
	\item show that, once auxiliary degrees of freedom are integrated out, it yields the original Hamiltonian [Eq.~(\ref{eq:extended_BH})) in the main text] with an effective pair-photon hopping that we compute;
	\item generalize the exact ground state of factorized many-body cat states found at $J = U/2$ to the full circuit (\textit{i.e.} including the auxiliary degrees of freedom);
	\item derive the second-quantized many-body Hamiltonian starting from a first-quantized  description of the underlying microscopic circuit;
	\item discuss the robustness of this implementation against likely mismatches in the fluxes.
\end{itemize}

\subsection{Second-quantized description of the circuit}
Our circuit is modeled by the following many-body Hamiltonian (the derivation from microscopic parameters is given in the next section)
\begin{multline}
	H=\sum_i \left\{\omega_{\rm{c}}a_{i}^{\dagger } a_{i}+\frac{U_0}{2}a_{i}^{\dagger 2} a_{i}^2\right\}+(2\omega_{\rm{c}}+\Delta)\sum_{\langle {i},{j}\rangle}b_{i,j}^\dagger b_{i,j}\\
	-\sum_{\langle {i},{j}\rangle}\left\{ \frac{\eta}{\sqrt{z}}\left[(a_{i}^{\dagger 2}+a_{j}^{\dagger 2}) b_{i,j}+(a_{i}^{2}+a_{j}^{2} ) b_{i,j}^\dagger\right]+ \frac{\eta^{(1)}}{\sqrt{z}}\left[(a_{i}^{\dagger}+a_{j}^\dagger) b_{i,j}+(a_{i}+a_{j}) b_{i,j}^\dagger\right]\right\}.\label{eq:BH_generalized}
\end{multline}
Here $b_{i,j}$ (resp. $b_{i,j}^\dagger$) is the annihilation (resp. creation) operator of the auxiliary resonator connecting the lattice sites $i$ and $j$, and $z$ is the number of nearest neighbors per site. The three-wave mixing term performs the coherent conversion of two qubit photons into a single photon of the auxiliary resonator. If $\Delta\gg U_0,\eta$, this conversion process is only virtual, and the single photon on the auxiliary resonator eventually goes back to the original lattice site, or is converted into two photons in the nearest-neighbor qubit. While the former process is expected to lead to a correction in the self-Kerr coupling of the lattice qubits, the latter process amounts to pair hopping: in this regime, the auxiliary resonators are expected to be mostly unoccupied and one can derive an effective Hamiltonian
\begin{figure}[t]
	\centering
	\includegraphics[width=0.8\columnwidth]{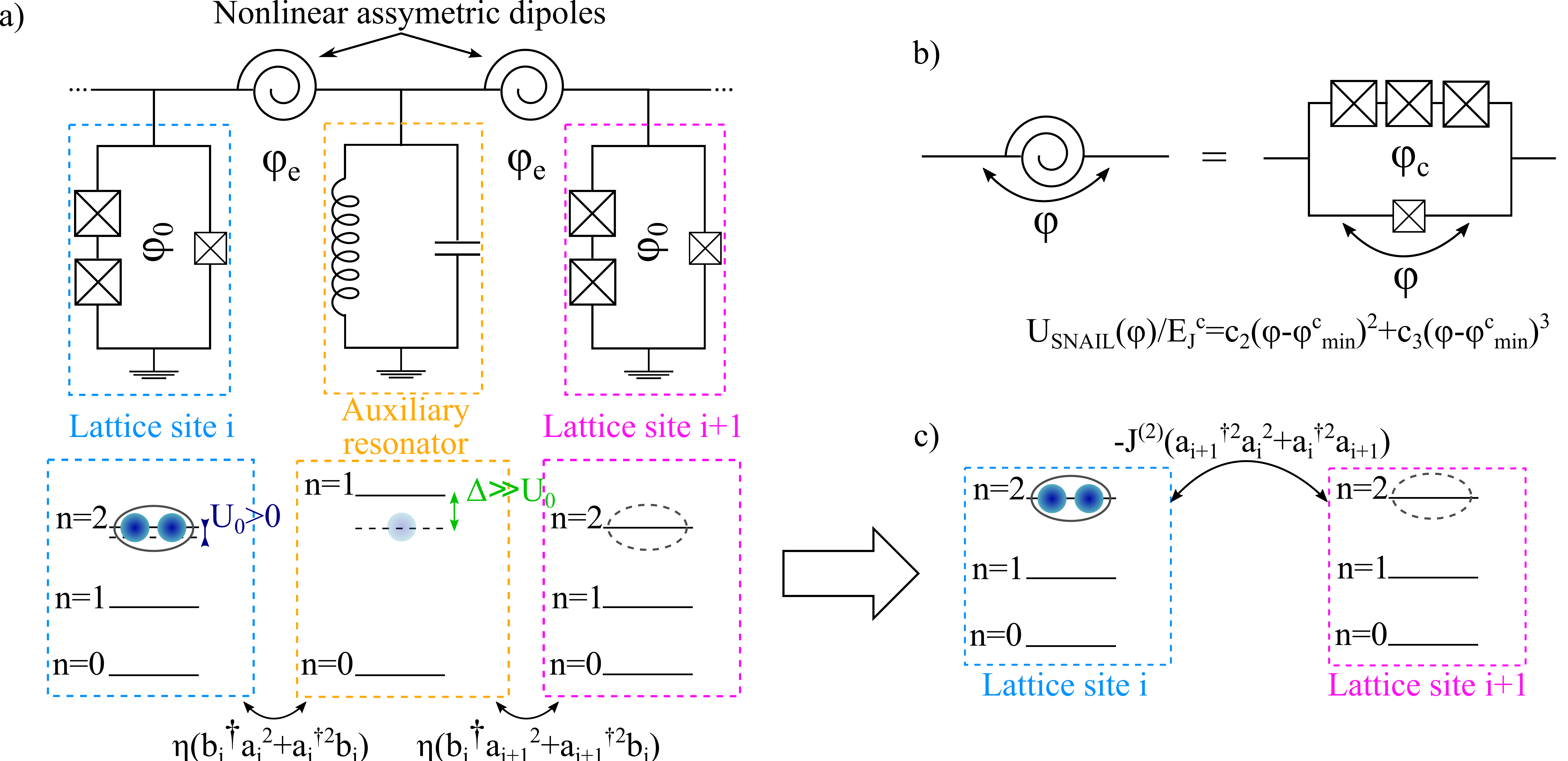} 
	\caption{\textbf{Proposal for implementation of two-photon hopping in a superconducting circuit architecture.}
		\textbf{(a)}: Description of the circuit and the emergent photon lattice model. Each lattice site is composed of a qubit with bare frequency $\omega_{\rm{c}}$. The qubit design, involving a small junction shunted with an array of two larger junctions with a flux $\varphi_0$, is inspired from the fluxonium qubit and is chosen in such a way to implement a Kerr nonlinearity $U$ with tunable strength and sign. In particular for $\varphi_0=\pi$, $U>0$ accounts for repulsive photon interactions. We suggest to couple the qubit to auxiliary resonators with a bare frequency $\omega_{\rm{aux}}=2\omega_{\rm{c}}+\Delta$ close to twice the lattice single-photon energy $\omega_{\rm{c}}$. \textbf{(b)}: Instead of a traditional Josephson junction-based nonlinear intersite coupling, the coupling to auxiliary resonators is implemented via three-wave mixing SNAIL (Superconduction Nonlinear Asymmetric Inductive eLement) devices~\cite{SNAIL,SNAIL_2} presenting a phase-dependent energy profile $U_{\rm{SNAIL}}(\varphi)=E_{\rm{J}}^c\left[c_2(\varphi-\varphi_{\rm{min}}^c)^2+c_3(\varphi-\varphi_{\rm{min}}^c)^3+\mathcal{O}(\varphi-\varphi_{\rm{min}}^c)^5\right]$. \textbf{(c)}: Effective lattice model emerging from the proposed circuit. At large detuning $\Delta\gg \eta,U$, pairs of photons can only be virtually converted to a single photon in the auxiliary resonator, leading to an emergent pair hopping between nearest qubits.
		\label{fig_supp:pair-hopping}}
\end{figure}
\begin{equation}
H_{\rm{eff}}=\omega_{\rm{c}}N+\frac{U_{\rm{eff}}}{2}\sum_i a_{i}^{\dagger 2} a_{i}^2- \frac{J}{z}\sum_{\langle {i},{j}\rangle}\left[a_{i}^{\dagger 2} a_{j}^2 + \mathrm{H.c.} \right]-\frac{J^{(1)}}{z}\sum_{\langle {i},{j}\rangle}\left[a_{i}^{\dagger} a_{j} + \mathrm{H.c.} \right]\,,\label{eq:BH_emergent}
\end{equation}
with a renormalized Kerr nonlinearity
\begin{equation}
U_{\rm{eff}}=U_0-\frac{2\eta^{2}}{\Delta}\label{eq_supp:effective_kerr}
\end{equation}
and a pair hopping constant
\begin{equation}
J=\frac{\eta^{2}}{\Delta}\,.\label{eq_supp:hopping_three_wave}
\end{equation}
For the sake of completeness, we also included single-photon hopping processes. However, due to the strong detuning $\omega_{\rm{aux}}-\omega_{\rm{c}}=\omega_{\rm{c}}+\Delta\simeq \omega_{\rm{c}}\gg|\Delta|,U_0$ these processes are far from resonance and are expected to be suppressed:
\begin{equation}
J^{(1)}=\frac{{\eta^{(1)}}^2}{\omega_{\rm{c}}}\,.\label{eq_supp:single-photon_hopping}
\end{equation}
In agreement with previous observations~\cite{SNAIL_2}, we find that three-wave mixing amplitudes as high as $\eta \approx 60\times 2\pi~\textrm{MHz}$ can be achieved with current technologies. However, the optimal regime to maximize the ratio $J/J^{(1)}$ between pair hopping and single-photon hopping was found for smaller values of $\eta$. As illustrated in Fig.~\ref{fig_supp:circuit_parameters}, using typical parameters of circuit QED, we find the following accessible ranges for the lattice parameters: 
\begin{eqnarray}
\eta& \approx &26\times 2\pi~\textrm{MHz}\\
J& \sim &0-10\times 2\pi~\textrm{MHz}\\
U_{\rm{eff}}&\sim&0-31\times 2\pi~\textrm{MHz}\\
J^{(1)}& \approx &675\times2\pi~\textrm{kHz}
\end{eqnarray}
The flux $\varphi_0$ was chosen intentionally to obtain a relatively small Kerr nonlinearity ($U_{0}\simeq 9 \times 2\pi~\textrm{MHz}$) so that pair hopping could compete with this effect. The particular ratio of interest, $J=U_{\rm{eff}}/2$, realized for this specific simulation at $\Delta=45\times 2\pi~\textrm{MHz}$, is accessible within our proposal. Moreover, we find that a ratio $J/J^{(1)} \approx 25$ can be achieved, which is far within the regime of stability of the pair-superfluid quantum phase against single-photon hopping (see Fig.~\ref{fig_supp:Gutzwiller_single_photon_hopping} in this Supplementary Material). This indicates that many-body cat states can be observed within this parameters range. For flux noise sensitive qubits, single-photon lifetimes $T=3~\mu\mathrm{s}$ corresponding to a loss rate $\gamma=300~1\times\textrm{kHz}$ are typically achieved experimentally, providing the model with the required time-scale separation between coherent and dissipative processes.

Let us note that one could modify the system connectivity in order to create exotic long-range pair hopping: by coupling all $N_{\rm{s}}$ qubits to a unique auxiliary resonator (via $N_{\rm{s}}$ different SNAILs), one could obtain an effective pair-hopping Hamiltonian where photons can tunnel by pair between arbitrary lattice sites, independently of the distance separating those sites. Using a resonator as a quantum bus has already been achieved experimentally for single-photon hopping~\cite{Quantum_bus}.

Finally, while the flux injection brings great tunability to the various couplings, this versatility can be traded off for a simpler architecture: for example, the flux qubits can be replaced by transmons with weaker anharmonicity (at the cost of having a negative $U_0$, which should not compromise the observation of many-body cat states), or by resonators with a self-Kerr (with tunable sign) provided by the coupling to an additional qubit.
\begin{figure}[t]
	\centering
	\includegraphics[width=0.4\columnwidth]{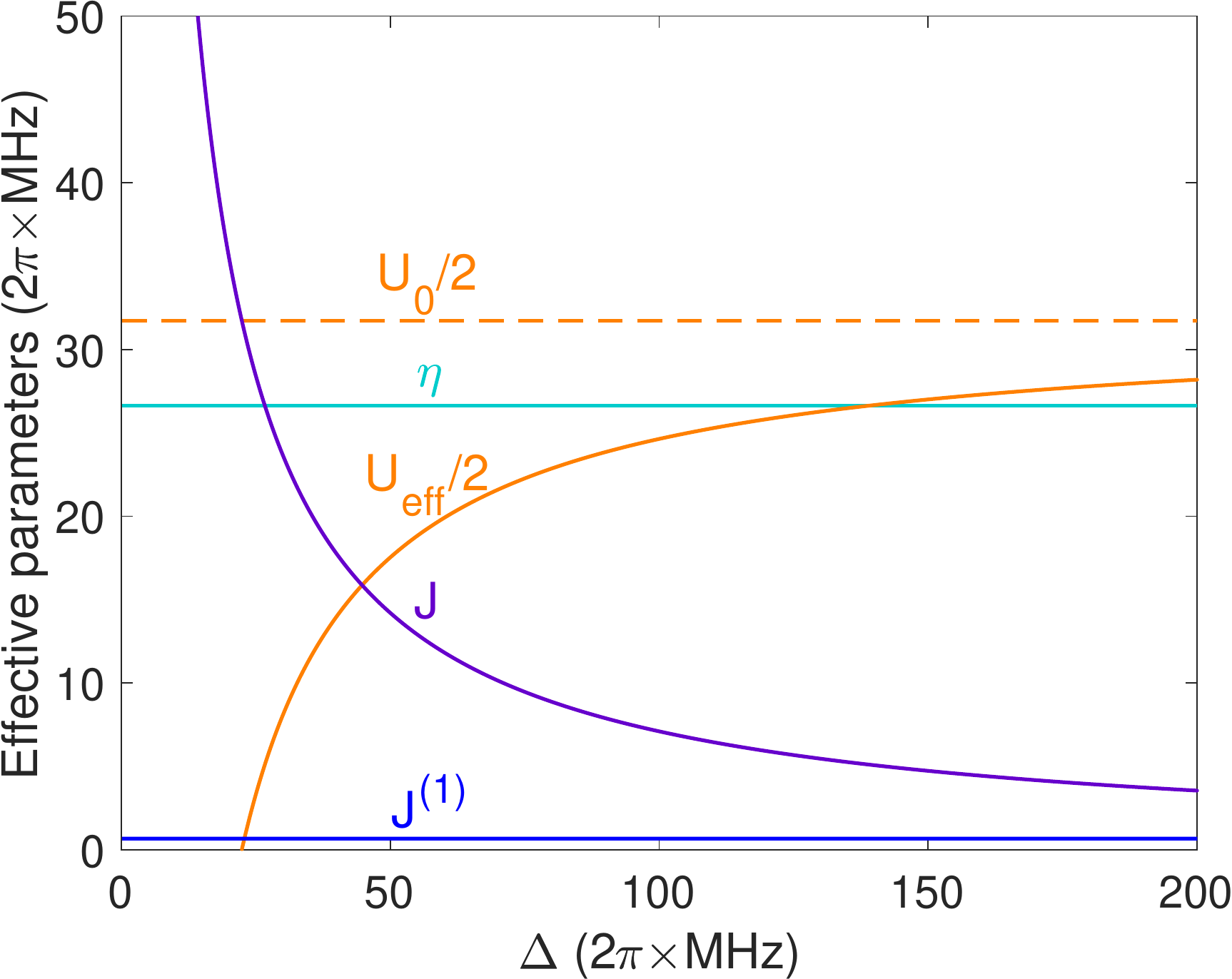} 
	\caption{\textbf{Predictions of model properties for typical circuit-QED parameters}. Three-wave mixing coefficient $\eta$, bare  (resp. renormalized) Kerr nonlinearity $U_0/2$ (resp. $U_{\rm{eff}}$), pair-hopping amplitude $J$, and single-photon hopping amplitude $J^{(1)}$  as a function of the detuning $\Delta$. Choice of circuit parameters: $E_J= 180 \times2\pi~\textrm{GHz}$ (large qubit junctions), $\alpha E_J= 35 \times2\pi~\textrm{GHz}$ (small qubit junction), $E_c=300\times2\pi~\textrm{MHz}$ (qubit charging energy), $\varphi_0=\pi$ (SNAIL flux), $E_J^c= 25\times2\pi~\textrm{GHz}$ (large SNAIL junction), $\alpha^{c}E_J^c=7.25\times2\pi~\textrm{GHz}$ (small SNAIL junction), $\varphi_c=0.92\pi$ (SNAIL flux), $Z=10\Omega$ (resonator impedance).
		\label{fig_supp:circuit_parameters}}
\end{figure}
\subsection{Generalized exact results}
The mapping between the three-waving mixing model in Eq.~(\ref{eq:BH_generalized}) and its pair-hopping counterpart in Eq.~(\ref{eq:BH_emergent}) only holds for significant detuning, $\Delta\gg\eta$, which constrains the experimentally accessible values of the pair hopping $J=\eta^{2}/\Delta$. Here, we show that upon neglecting single-photon hopping, the three-wave mixing model in Eq.~(\ref{eq:BH_generalized}) is interesting on its own, and exact results related to many-body cat states can be derived even when the condition $\Delta\gg\eta$ is not satisfied. 

As a first step, we move to the rotating frame: $a_i\to a_i \textrm{exp}(\rmi\omega_{\rm{c}}t)$, $b_{i,j}\to b_{i,j} \textrm{exp}(2\rmi\omega_{\rm{c}}t)$, shifting the bare photon frequency by $-\omega_{\rm{c}}$. The Hamiltonian in Eq.~(\ref{eq:BH_generalized}) becomes:
\begin{equation}
\tilde{H}=\frac{U_0}{2}\sum_ia_{i}^{\dagger 2} a_{i}^2+\Delta\sum_{\langle {i},{j}\rangle}b_{i,j}^\dagger b_{i,j}
-\frac{\eta}{\sqrt{z}}\sum_{\langle {i},{j}\rangle} \left[(a_{i}^{\dagger 2}+a_{j}^{\dagger 2}) b_{i,j}+(a_{i}^{2}+a_{j}^{2} ) b_{i,j}^\dagger\right]\,. \label{eq:BH_generalized_2}
\end{equation}
For a magic value $\eta=\eta^{\ast}\equiv\sqrt{U_0\Delta}/2$ of the three-wave coupling, one can show that the generalized states
\begin{equation}
\ket{\psi^P(\alpha)}_{\rm{gen}}=\bigotimes_{i}\ket{\mathcal{C}^{P(i)}(\alpha)}_{a_i}\bigotimes_{\langle i,j\rangle}\ket{\beta}_{b_{i,j}},\label{eq:supp_generalized_cat}
\end{equation}
composed of a product of cat states on the main lattice qubits and of coherent states on the auxiliary resonators, are exact zero-energy eigenstates of the Hamiltonian in Eq.~(\ref{eq:BH_generalized_2}). The amplitude $\beta=2\eta\alpha^2/(\Delta\sqrt{z})$ is conditioned by the amplitude $\alpha$, which can take arbitrary values, and the qubit local parities $P(i)$ can vary from site to site. 

Similarly to the case discussed in the main manuscript, in the original non-rotating frame one needs to add a contribution  $\omega_{\rm{c}}(\sum_i a_{i}^\dagger a_i+2\sum_{\langle i,j\rangle} b_{i,j}^\dagger b_{i,j})$ to the Hamiltonian $\tilde{H}$ in order to recover the original Hamiltonian $H$ of Eq.~(\ref{eq:BH_generalized}). In this case, while the states in Eq.~(\ref{eq:supp_generalized_cat}) are no longer eigenstates, they follow the simple and exact equation of motion: $\ket{\psi(t)}=\ket{\psi^P(\alpha\, \textrm{exp}(-\rmi\omega_{\rm{c}}t))}_{\rm{gen}}$. Moreover, exact eigenstates of the full Hamiltonian $H$ can be obtained by projecting the states in Eq.~(\ref{eq:supp_generalized_cat}) on a subspace with a conserved total number of excitations $N_{\rm{gen}}=\sum_i a_{i}^\dagger a_i+2\sum_{\langle i,j\rangle} b_{i,j}^\dagger b_{i,j}$:
\begin{equation}
\ket{\psi^P_N}_{\textrm{gen}}=\mathcal{P}_{\textrm{gen},N}\left(\ket{\psi^P(\alpha)}_{\textrm{gen}}\right)\,.
\end{equation}
Let us note that once Eqs.~(\ref{eq_supp:effective_kerr}) and~(\ref{eq_supp:hopping_three_wave}) are injected into the expression of $\eta_c$, the condition $\eta=\eta_c $ is equivalent to $J=U_{\rm{eff}}/2$, in agreement with the derivation of the pair-hopping model and the results of the main manuscript. The novelty comes from the fact that this relation extends for a small detuning $\Delta\leq \eta$, where the auxiliary resonators are significantly populated. Ultimately, the three-wave mixing model in Eq.~(\ref{eq:BH_generalized}) offers a even more general and flexible framework for the study of many-body cat states compared to the pair-hopping model that we study in the main text.

\subsection{Derivation from microscopic circuit parameters}
We now present the microscopic derivation of the lattice parameters of the Hamiltonian in Eq.~(\ref{eq:BH_generalized}) of the Supplementary Material, starting from the circuit proposal in Fig.~(\ref{fig_supp:pair-hopping}).
The circuit Hamiltonian can be written as a function of the number and phase conjugated variables $n_i$ and $\varphi_i$ associated to the qubits, and the number and phase variables $n^{\textrm{aux}}_{i,j}$ and $\varphi^{\textrm{aux}}_{i,j}$  associated to the auxiliary resonators as
\begin{equation}
H=\sum_i H_i+\sum_{i,j}H^{\textrm{aux}}_{i,j}+H_{\textrm{SNAIL},i,j}
\end{equation}
where
\begin{equation}
\label{eq:Fluxonium}
H_i=4 E_c n_i ^2-E_J\left[2\textrm{cos}(\varphi_i/2+t\varphi_e)-\alpha\textrm{cos}(\varphi_i+\varphi_0+t\varphi_e)\right] 
\end{equation}
is the Hamiltonian of the lattice qubits ,
\begin{equation}
H^{\textrm{aux}}_{i,j}=4 E_c^{\rm{aux}} n^{\textrm{aux}2}_{i,j} +E_J^{\rm{aux}} \varphi^{\textrm{aux}2}_{i,j}/2
\end{equation}
is the Hamiltonian of the auxiliary resonators, and
\begin{equation}
H_{\textrm{SNAIL},i,j}=E_{\textrm{SNAIL},i,j}(\varphi_i-\varphi^{\textrm{aux}}_{i,j}+(1-t)\varphi_e)+E_{\textrm{SNAIL},i,j}(\varphi_j-\varphi^{\textrm{aux}}_{i,j}+(1-t)\varphi_e),
\end{equation}
is the Hamiltonian of the SNAIL devices. As shown in Ref.~\cite{SNAIL}, by injecting a well-chosen flux $\varphi_c$ in the loop of the SNAIL device
its inductive energy landscape
$$
E_{\textrm{SNAIL},i,j}(\varphi)=E_J^{c}\left[c_2\left(\varphi-\varphi_{\rm{min}}^c\right)^2+c_3\left(\varphi-\varphi_{\rm{min}}^c\right)^3+\mathcal{O}\left(\varphi-\varphi_{\rm{min}}^c\right)^5\right]
$$
is minimal around a non-trivial phase $\varphi_{\rm{min}}^c$, and possesses a third order nonlinearity but no fourth order term. The SNAIL device is assumed to be shunted by a large capacitor, and its charging energy was not included (although a non-zero small charging energy would only induce small modifications of the coupling values of our effective in our model). The Josephson energy $E_J^{c}$ is assumed to be small with respect to $E_J$, so the SNAIL is essentially seen as a coupler between the qubit and resonator modes. The system is assumed to be configured in a `transmonic' regime $E_J,E_J^c,E_J^{\rm{aux}}\gg E_c,E_c^{\rm{aux}}$ where phase fluctuations are strongly suppressed: the system properties associated to quantum fluctuations can then be obtained by expanding the Hamiltonian $H$ in powers of $\varphi_i$, $\varphi^{\textrm{aux}}_{i,j}$ around its classical minimum and using the standard commutation rule $[\varphi_i,n_i]= \rmi\hbar$, thus neglecting  the possibility of a phase winding. 

An additional flux $\varphi_e$ is injected in the loops containing nearest neighboring qubits and auxiliary resonators. As we will see, this flux plays a role in minimizing separately the inductive energies of the qubit and SNAIL devices. The dimensionless parameter $t$ is nonphysical and can take arbitrary values by gauge transformation (we specify below our choice of gauge). In the specific case of a flux bias $\varphi_0=\pi$, once the nonphysical gauge variable $t$ is set to zero, the qubit Hamiltonian in Eq.~(\ref{eq:Fluxonium}) has an extremum around $\varphi_i=0$, yielding:
\begin{equation}
H_i=E_0+\left\{4 E_c n_i ^2+E_J\left[\left(\frac{1}{2}-\alpha\right)\frac{\varphi_i^2}{2}+\left(\alpha-\frac{1}{8}\right)\frac{\varphi_i^4}{4!}\right]\right\}.
\end{equation}
Thus, for $1/8<\alpha<1/2$ and in absence of coupling to the rest of the circuit, the qubit is stable around $\varphi_i=0$ and has  a repulsive Kerr nonlinearity, in stark contrast with a standard transmon, which is always attractive. Moreover, for $\varphi_0=0$ the qubit has an attractive Kerr. Let us note that this regime has a specific value of $\varphi_0$ for which the Kerr constant exactly cancels out (this  is actually the physical mechanism underlying the engineering of the purely three-wave mixing SNAIL device). Therefore, the Kerr constant is tunable in sign and strength via the in loop flux $\varphi_0$. The coupling of the qubit to the rest of the circuit will only slightly affect the range of values of $\alpha$ for which these features are valid.

In the case of arbitrary $\varphi_0$ and for  $\varphi_e=0$, the minimum of the inductive part of $H_i$ is not generally located around $\varphi_i=0$ (the same is true for SNAIL devices). This can be corrected by inserting a non-zero flux bias $\varphi_e$  in the loops containing nearest neighboring qubits and auxiliary resonators. We now fix the gauge choice by choosing the free variable $t$ such that $\varphi_{\rm{min}}=t\varphi_e$, $\varphi_{\rm{min}}^c=(1-t)\varphi_e$: both the minimum of inductive energy of the qubits and of the SNAIL are then simultaneously realized  at $\varphi_i=0$, $\varphi_{\textrm{aux},i,j=0}$, and around this minimum one has
\begin{equation}
\label{eq:Fluxonium_low}
H_i=E_0+\left\{4 E_c n_i ^2+E_J^{\rm{min}}\frac{\varphi_i^2}{2}+\epsilon_J^{\rm{min}}\frac{\varphi_i^4}{4!}\right\}+\beta E_J\frac{\varphi_i^3}{3!}.
\end{equation}
where $E_J^{\rm{min}}=E_J\left(\textrm{cos}(\varphi^{\rm{min}}/2)/2+\alpha\textrm{cos}(\varphi^{\rm{min}}+\varphi_0)\right)$, $\beta=\textrm{sin}(\varphi^{\rm{min}}/2)/4+\alpha\textrm{sin}(\varphi^{\rm{min}}+\varphi_0)$, and 
\begin{equation}
H_{\textrm{SNAIL},i,j}=E_J^c\left\{ c_{2}\left[\varphi_i^2+\varphi_j^2+\varphi^{\textrm{aux}2}_{i,j}-2(\varphi_i\varphi^{\textrm{aux}}_{i,j}+\varphi_j\varphi^{\textrm{aux}}_{i,j})\right]+c_3\left[(\varphi_i-\varphi^{\textrm{aux}}_{i,j})^3+(\varphi_i-\varphi^{\textrm{aux}}_{i,j})^3\right]\right\}.
\end{equation}
The effective model in Eq.~(\ref{eq:BH_generalized}) is obtained via a rather standard procedure: we first study the quadratic and local part of the Hamiltonian 
\begin{eqnarray}
H_{\rm{loc}}^{(2)}&=&\sum_{i}\left[4 E_c n_i ^2+\frac{E_J^{\rm{eff}}}{2}\varphi_i^2\right]+\sum_{i,j}\left[4 E_c^{\rm{aux}} n^{\textrm{aux}2}_{i,j} +\frac{E_J^{\rm{aux,eff}}}{2} \varphi^{\textrm{aux}2}_{i,j}\right]\\
&=&C+\omega_{\rm{c}}\sum_{i}a_i^\dagger a_i+\omega_{\rm{aux}}\sum_{i,j}b_{i,j} b_{i,j},
\end{eqnarray}
where $E_J^{\rm{eff}}=E_J^{\rm{min}}+2c_2E_J^c $ (resp. $E_{J}^{\textrm{aux,eff}}=E_J^{\rm{min}}+2c_2 E_J^c $), and
\begin{eqnarray}
a_i&=&\frac{1}{2}\left(\frac{\varphi_i}{\varphi_{\rm{ZPF}}}+\textrm{i} \frac{n_i}{n_{\rm{ZPF}}}\right)\\
b_{i,j}&=&\frac{1}{2}\left(\frac{\varphi_{i,j}^{\rm{aux}}}{\varphi_{\rm{ZPF}}}+\textrm{i}\frac{n_{i,j}^{\rm{aux}}}{n_{\rm{ZPF}}}\right)
\end{eqnarray}
are the annihilation operators on the i-th qubit and the auxiliary resonator between the i-th and j-th sites. The quantities
\begin{eqnarray}
\varphi_{\rm{ZPF}}&=&(2E_c/E_J^{\rm{eff}})^{1/4}\\
\varphi^{\rm{aux}}_{\rm{ZPF}}&=&\varphi^{\rm{aux}}_{\rm{ZPF}}=(2E_c^{\rm{aux}}/E_J^{\rm{aux,eff}})^{1/4}
\end{eqnarray}
amount respectively to the phase fluctuations in the qubit and auxiliary resonator in the ground-state of the local and quadratic Hamiltonian $H_{\rm{loc}}$. The Cooper pairs number fluctuations $n_{\rm{ZPF}}=1/(2\varphi_{\rm{ZPF}})$ [resp. $n_{\rm{ZPF}}^{\rm{aux}}=1/(2\varphi_{\rm{ZPF}}^{\rm{aux}})$] are inversely related to the phase fluctuations. The bare frequency  of the qubits and the resonators are given by 
\begin{eqnarray}
\omega_{\rm{c}}&=&\sqrt{8 E_c E_{\rm{J}}^{\rm{eff}}}+E_c\epsilon_J^{\rm{min}}/E_J^{\rm{eff}}\\
\omega_{\rm{aux}}&=&\sqrt{8 E_c^{\rm{aux}} E_{\rm{J}}^{\rm{aux,eff}}}.
\end{eqnarray}

Injecting the parameters of Fig.~(\ref{fig_supp:circuit_parameters}), we find $E_J^{\rm{eff}}=59\times2\pi~\textrm{GHz}\gg E_c=\textrm{300}\times 2\pi~\textrm{MHz}$: the qubit is effectively in the transmonic regime, and its zero-point phase fluctuations $\varphi_{\rm{ZPF}}=0.32$  are relatively weak with respect to $2\pi$. With the choice of impedance $Z=10\Omega=4\pi R_{Q}$, where $R_Q=(h/2e)^2$ is the resistance quantum, we obtain $\varphi^{\rm{aux}}_{\rm{ZPF}}=(\pi Z/R_Q)^{1/2}=0.07$: the auxiliary resonator is even more strongly `transmonic'. In this regime, since $\varphi_i=\varphi_{\rm{ZPF}}(a_i+a_i^{\dagger})$, higher powers of the phase variable bring smaller contributions to the Hamiltonian. Likewise, since we assumed $E_{J},E_{J}^{\rm{aux}}\gg E_J^c$, the non-local quadratic part of the Hamiltonian is treated perturbatively with respect to the local quadratic part. In this framework, all non-local and nonlinear terms  that do not preserve the photon number can be safely neglected, except for the cubic term  $\propto(\varphi_i^2+\varphi_j^2) \varphi^{\textrm{aux}}_{i,j}$ which brings a contribution  $\propto[(a_i^2+a_j^2) a_{i,j}^{\dagger}+\textrm{H.c.}]$ (we want to operate in a regime where $\omega_{\rm{aux}}$ is close to $2\omega_{\rm{c}}$). Keeping relevant terms, in the rotating wave approximation one obtains the Hamiltonian in Eq.~(\ref{eq:BH_generalized}) with 
\begin{eqnarray}
\eta &=&-3\sqrt{z} c_3 E_J^c \varphi_{\rm{ZPF}}^2 \varphi_{\rm{ZPF}}^{\rm{aux}}\\
\eta^{(1)} &=& -2 \sqrt{z} c_2 E_J^c \varphi_{\rm{ZPF}}\varphi_{\rm{ZPF}}^{\rm{aux}}\\
U_0 &=&\frac{\epsilon_J^{\rm{min}}}{E_{J}^{\rm{eff}}}E_c
\end{eqnarray}
where $\epsilon_J^{\rm{min}}\propto E_J$ is strength of the quartic term in the expansion in Eq.~(\ref{eq:Fluxonium_low}) of the qubit inductive Hamiltonian around its energy minimum.

\subsection{Robustness against flux mismatch}
An experimental implementation of this proposal  relies on the ability to control individually the values of many fluxes. 
In practice, the various fluxes are typically controlled via a unique magnetic field, and their respective values are set by the respective loop cross sections. Therefore, individual fine flux-tuning is hardly achievable in the post-engineering phase. Here, we discuss the robustness of our proposal against flux mismatch: we argue that the main properties  of our model survive even under `bad' flux choices as long as the SNAIL devices, resonators and qubit are reproducible with good enough accuracy. 

First, we stress that the main ingredients of our model are still present for generic fluxes as long as the resonator frequency is close to twice the qubit frequency (which implies that single-photon hopping is strongly suppressed), and that the SNAIL performs three-wave mixing (implying pair hopping). The latter property is verified as long as $\varphi_c \neq 0 [\pi]$. The ability to reduce and eventually to cancel the self-Kerr of the qubit via the unique flux knob guarantees that this pair hopping can be made large enough to compete against interactions and ultimately reach the special value $J_*=U_{\rm{eff}}/2$. 

The main consequence of a flux mismatch is to introduce four-wave mixing in the SNAIL device. This is responsible for a self-Kerr term $\sum_{i,j}\frac{U^{\rm{aux}}}{2}b_{i,j}^{\dagger 2} b_{i,j}^2$ in the resonator and a cross-Kerr term $\Delta H_{\rm{cross-Kerr}}=\sum_{i,j}V(a_i^\dagger a_i +a_j^\dagger a_j )b_{i,j}^\dagger b_{i,j}$ between the resonators and the qubit. However, the various circuit elements being in the transmonic regime, four-wave mixing processes are typically weaker than three-wave mixing processes. In particular, due to the weak resonator impedance value $Z=10\Omega\lll R_{Q}\simeq 6.2\, \textrm{k}\Omega$, the cross-Kerr coupling and  the self-Kerr amplitude of the resonator are strongly suppressed: injecting the circuit parameters of Fig.~\ref{fig_supp:circuit_parameters}, we found in fact the upper bound $V_{\rm{max}}=3.6\times 2\pi~\textrm{MHz}$, and $U^{\rm{aux}}_{\rm{max}}=200 \times 2\pi~\textrm{kHz}$ for the cross-Kerr and the resonator self-Kerr amplitudes in the worst case scenario, \textit{i.e.} when a bad choice of flux maximizes four-wave mixing processes within the SNAIL. These represent small perturbations compared to the energy scale of three-wave mixing processes $\eta\simeq 26 \times 2\pi~\textrm{MHz}$. One concludes that the three-wave model Eq.~(\ref{eq:BH_generalized}) and the emergent pair-hopping model Eq.~(\ref{eq:BH_emergent}) should both be robust against a flux mismatch, and that its impact is limited to a renormalization of the lattice parameters $\eta$, $U_0$, $\Delta$, $U_{\rm{eff}}$, $J$.

\section*{\uppercase{Gutzwiller Mean-Field analysis}} The Gutzwiller Mean-Field method is a well-established numerical technique consisting in approximating the many-body quantum state by a state which is factorized over the various lattice sites. It is exact in the limit $z\to +\infty$ of a large number of nearest neighbors per lattice site, \textit{i.e.}, for a lattice with infinite spatial dimensionality, or in the case of a long-range hopping. At finite $z$, it usually captures important aspects of the real phase diagram.

For a zero-temperature equilibrium situation, the Gutzwiller ansatz is implemented in a variational fashion where the ground-state wave function is assumed to take the form $\ket{\psi_{0}}=\bigotimes_i \ket{\psi_{i}}$, and one minimizes the average energy computed with $H_0-\mu N$ in order to find the optimal $\ket{\psi_{0}}$. While this procedure can be easily adapted to describe the spontaneous breaking of the spatial translation symmetry, we focus on a homogeneous ansatz $\ket{\psi_{i}}=\ket{\psi_{\rm{GW}}}$ for all $i$. Within the considered Bose-Hubbard model with pair-hopping, this leads to the minimization of the quantity
\begin{equation}
\langle \psi_{\rm{GW}}|-\mu a^\dagger a+\frac{U}{2}a^{\dagger 2}a^2|\psi_{\rm{GW}}\rangle-J \left|\langle \psi_{\rm{GW}}|a^2|\psi_{\rm{GW}}\rangle\right|^2
\end{equation}
on the single-site wave-function $\ket{\psi_{\rm{GW}}}$.

Correspondingly, in the driven-dissipative situation one uses a factorized ansatz for the density matrix of the system composed of the photons and the two-level emitters, $\rho_{0}=\bigotimes_i \rho_{i}$. We also assume translational invariance: $\rho_{i}=\rho_{\rm{GW}}$. By inserting this ansatz in the master Eq.~(\ref{eq:master_equation})) (main text), and by keeping only the most relevant terms in $1/z$, we get an effective master equation
\begin{equation}
\partial_{t}\rho_{\rm{GW}}=- \rmi \com{H_{\rm{GW}}(t)}{\rho_{\rm{GW}}}+\mathcal{L}(\rho_{\rm{GW}}),\label{eq_lorentz:ev_Gutzwiller}
\end{equation}
for the single-site density matrix $\rho_{\rm{GW}}$. The Gutzwiller Hamiltonian $H_{\rm{GW}}(t)=H_{\rm{loc}}+H_{\rm{tun}}(t)$ is the sum of the local contributions of the photon-emitter Hamiltonian
\begin{equation}
H_{\rm{loc}}=\omega_{\rm{c}}a^{\dagger}a+\omega_{\rm{at}}\sigma^{+}\sigma^{-}+\Omega_{\rm{R}}\left(a^{\dagger 2}\sigma^-+a^2\sigma_i^+\right)
\end{equation}
and of a time-dependent Mean-Field term
\begin{equation}
H_{\rm{tun}}(t)=-J\left[\psi(t)a^{\dagger 2}+\psi(t)^*a^{2}\right]
\end{equation}
which corresponds to non-local hopping processes and has to be computed dynamically and self-consistently using $\psi(t) = \left\langle a^2\right\rangle (t)=\textrm{Tr}\left[\rho_{\rm{GW}}(t)a^2\right]$. Finally, $\mathcal{L}(\rho_{\rm{GW}})=\Gamma_{\rm{l}}\,\,\mathcal{D}[a^2](\rho_{\rm{GW}})+\Gamma_{\rm{p}}\,\,\mathcal{D}[\sigma^+](\rho_{\rm{GW}})$ is the sum of all local dissipative processes. The translational invariance justifies use the simplified notations $a_i\to a$, $\sigma^{-}_i\to \sigma^{-}$. 

To compute the steady state in the even parity sector, we initialize the system in a large cat state $\ket{\psi_{\rm{GW}}}=\ket{\mathcal{C}^{+}(\alpha)}$ and let it evolve under the dynamics provided by Eq.~(\ref{eq_lorentz:ev_Gutzwiller}) until the system reaches a stable configuration. In particular, we found a steady-state oscillatory order parameter $\psi^{(2)}(t)=\psi_0^{(2)}\rme^{-\rmi\omega_{\rm PSF}t}$. Wherever we find $\psi^{(2)}_0=0$ we conclude to the presence of a normal phase, while non-trivial solutions $\psi^{(2)}_0\neq 0$ correspond to the existence of pair superfluidity. Only one stable solution was found  for a choice of cavity frequency verifying $\omega_1\leq 2\omega_c\leq\omega_2$, where $\omega_1$ and $\omega_2$ are the two frequencies for which the pump spectrum $\mathcal{S}_{\rm{em}}(\omega)$ given in Eq.~(7) (main text) compensates the losses: $\mathcal{S}_{\rm{em}}(\omega_{1/2})=\Gamma_{\rm{l}}$. Outside this interval, two stable solutions $\psi_0^{(2)}$ were found. The solution that is reached in the steady state depends on the choice of $\alpha$ for the initial conditions. For the sake of simplicity, in Fig.~\ref{fig:Gutzwiller_dissipative}c of the main text we only represent the solution with the maximal $|\psi_0^{(2)}|$.
\section*{\uppercase{Proof of exact results on the ground-states and thermodynamic stability}}
In this Supplementary Material, we detail the proof that the states of Eq.~(3) in the main text are indeed the ground-states of $H_0$ at $J=U/2$, and we compute exactly the stability domain of $H_0-\mu N$. 

First, applying the Hamiltonian $H_0$ to $\ket{\psi^{P}(\alpha)}$ yields
\begin{equation}
H_{0}\ket{\psi^{P}(\alpha)}=\sum_{i}a_{i}^{\dagger 2}\left[\frac{U}{2}-J\right]\alpha^2\ket{\psi^{P}(\alpha)}\,,
\end{equation}
which vanishes at $J_\ast=U/2$, thus establishing that $\ket{\psi^{P}(\alpha)}$ is a zero-energy eigenstate of $H_{0}$. Second, $H_{0}$ is a positive operator for $J= U/2$. We will demonstrate this properties more generally for $J\leq U/2$. To proceed, we consider the Fourier transforms $B_{k}=1/\sqrt{N_{\rm sites}} \, \sum_j \exp{( \rmi k\cdot j)}a_j^2$ of the operators $a_j^2$ annihilating pairs of photons. Here, $k$ is a momentum vector  in the Brillouin zone (BZ) of the reciprocal lattice. It is then possible to rewrite the Hamiltonian in terms of the $B_k$ operators
\begin{equation}
\label{eq:hamiltonian-momentum}
H_{0}=\sum_{k\in \rm{BZ}} \tilde{\epsilon}_k B_{k}^\dagger B_k,
\end{equation} 
where $\tilde{\epsilon}_k\equiv U/2-J/z\sum_{d} \textrm{cos}(k\cdot d)$, and the vectors $d$ represent all the possible displacements towards the $z$ nearest neighbors of a given lattice site. When $J\leq U/2$, one has $\tilde{\epsilon}_k\geq 0$ for all $k$: $H_0$ is thus the sum of positive definite matrices $\tilde{\epsilon}_k B_{k}^\dagger B_k$ and is thus as well positive. 

This argument can be easily adapted to demonstrate that $J <U/2$ is the stability domain in the case of the grand-canonical ensemble: given that $U/2-J\leq\tilde{\epsilon}_k$ for all momenta $k$,  the spectrum of the grand-canonical Hamiltonian $H_0-\mu N$ is necessarily bounded from below if the spectrum of $H_{\rm low}=-\mu N+(U/2-J)\sum_{k\in \rm{BZ}} B_{k}^\dagger B_k$ possesses a lower bound as well. Back to the real-space representation $a_i$, one can see that  $H_{\rm low}=-\mu N+(U/2-J)\sum_i a_i^{\dagger 2}a_i^2$ is the Bose-Hubbard Hamiltonian with a strictly repulsive attraction $U/2-J>0$ and a vanishing hopping. This Hamiltonian is known to be thermodynamically stable for any $\mu$, thus completing the proof. For $J> U/2$, using a coherent state $\bigotimes_i\ket{\alpha}$ as a variational ansatz yields a Mexican hat-shaped energy landscape $-\mu |\alpha|^2 +\left(U/2-J\right)|\alpha|^4$ unbounded from below, indicating the presence of a thermodynamic instability.
\section*{\uppercase{Gutzwiller with even and odd parities}}
In this section, we discuss the zero-temperature phase diagram of the two-photon hopping Bose-Hubbard model when now including both even and odd on-site parity sectors. We argue that this does not modify significantly the phenomenology that was exposed in the main manuscript where only the even parity sector was included. In particular, the existence of a pair-superfluid phase is still expected since the two sectors are not coupled by the Hamiltonian dynamics. This ensures that the ground state may still break $U(1)$ while preserving $\mathbb{Z}_2$.

We used a Gutzwiller (mean-field) approach to compute the phase diagram presented in Fig.~\ref{fig_supp:Gutzwiller_parities}a.  
The main addition to the one presented in Fig.~\ref{fig:Gutzwiller_equilibrium}a of the main manuscript consists in the presence of Mott regions at every integer density, $n=0,1,2,3 ...$, rather than at even integer density only. The Mott regions no longer take the shape of lobes closing and meeting at $J=0$, but they are now separated by first-order phase transitions (where local observables such as the density are discontinuous) and the Mott-to-superfluid phase transition always takes place at finite pair hopping $J > 0$. Indeed, pair hopping does not allow the mobility of single doublon or hole excitations, and only on-site doublon/hole pair excitations have mobility for $J\ll U$. 
Besides these modifications, we recover most of the results obtained when truncating to the even parity sector. In particular, the overlap with a cat state still approaches unity as $J$ approaches $U/2$.



We completed the analysis by  computing the many-body energy gap $\Delta=|E_{\rm{ev}}-E_{\rm{odd}}|/N_{\rm{sites}}$ separating the ground states $\ket{\mathrm{GS}_{\rm{ev}}}$ and $\ket{\mathrm{GS}_{\rm{odd}}}$ with either even or odd parities at all sites, and we found it to be exponentially suppressed as $J$ approaches $U/2$ (see Fig.~\ref{fig_supp:Gutzwiller_parities}d):
\begin{equation}
\label{eq_supp:scaling_gap}
\Delta \underset{J\to U/2}{\sim} \rme^{- \frac{\mathcal{A}}{U - 2J}}\,,
\end{equation}
where $\mathcal{A} > 0$. Considering that $\psi^{(2)}\sim \mu/(U-2J)$ close to the instability at $J=U/2$, the scaling in Eq.~(\ref{eq_supp:scaling_gap}) is equivalent to an exponential suppression $\Delta\propto \exp (- C\psi^{(2)})$ of the gap in the PSF order parameter.

\begin{figure}[t]
	\centering
	\includegraphics[width=0.6\columnwidth]{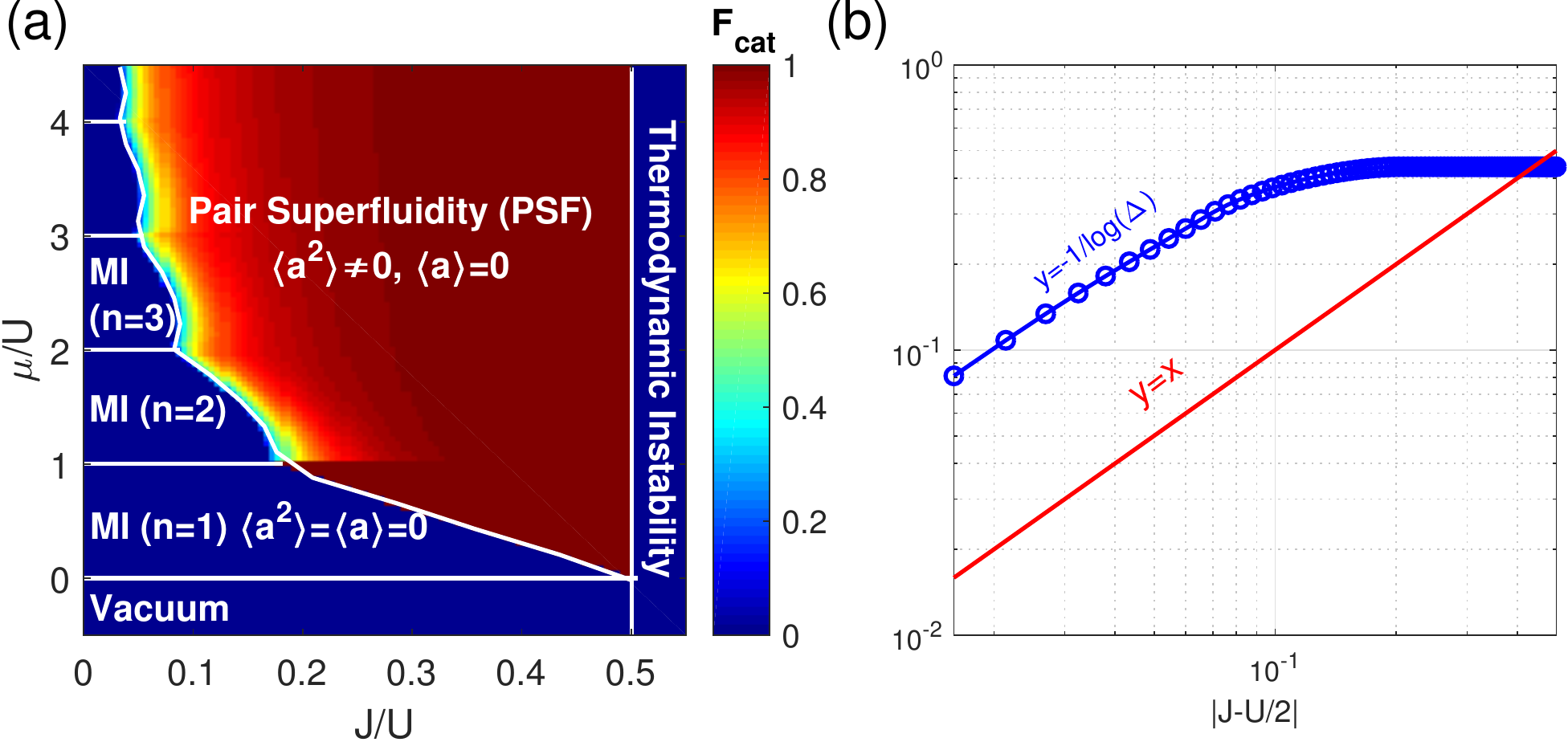} 
	\caption{\textbf{(a)}: Zero-temperature phase diagram of the Bose-Hubbard model with two-photon hopping, including both even and odd parity sectors. The fidelity $F_{\rm{cat}}=\textrm{max}_{\alpha,\pm}|\langle\mathcal{C}^{\pm}(\alpha)|\psi_{\rm{GW}}\rangle$ of the Gutzwiller wave function with a cat state is represented in color plot. \textbf{(b)}: Dependence on $J$ of the many-body energy gap $\Delta=|E_{\rm{ev}}-E_ {\rm{odd}}|/N_{\rm{sites}}$ separating the even and odd parity sectors, see Equation~(\ref{eq_supp:scaling_gap}) of this Supplementary Material.
		\label{fig_supp:Gutzwiller_parities}}
\end{figure}
\section*{\uppercase{Robustness of pair superfluidity against single-particle hopping}}
The Gutzwiller analysis presented above revealed a small but finite many-body energy gap $\Delta=|E_{\rm{ev}}-E_{\rm{odd}}|/N_{\rm{sites}}$ separating the ground states $\ket{\mathrm{GS}_{\rm{ev}}}$ and $\ket{\mathrm{GS}_{\rm{odd}}}$ with either even or odd parities at all sites. The finiteness of this gap for $J<U/2$ suggests that pair superfluidity is not only an accidental consequence of the local $\mathbb{Z}_{2}^{\rm{loc}}$ symmetry, but is also expected to be robust against small but finite symmetry-breaking perturbations coupling the two parity sectors. In order to test this claim, we considered an extended Bose-Hubbard Hamiltonian
\begin{equation}
H_{1}=-\mu N+\frac{U}{2}\sum_i a_{i}^{\dagger 2} a_{i}^2- \frac{J}{z}\sum_{\langle {i},{j}\rangle}\left[a_{i}^{\dagger 2} a_{j}^2 + \mathrm{H.c.} \right]-\frac{J^{(1)}}{z}\sum_{\langle {i},{j}\rangle}\left[a_{i}^{\dagger} a_{j} + \mathrm{H.c.} \right]\,,
\end{equation}
where a single-particle hopping term with hopping amplitude $J^{(1)}$ has been included in addition to the pair hopping.

The zero-temperature phase diagram of the above Hamiltonian has been computed by means of a Gutzwiller mean-field approach and is presented in Fig.~\ref{fig_supp:Gutzwiller_single_photon_hopping}. To understand the competition between single-particle- and pair-hopping processes, we have determined the various phases as a function of $J/U$ and $J^{(1)}/U$ at fixed chemical potential $\mu/U=1.5$ by monitoring both order parameters $\psi^{(1)} \equiv \langle a_i \rangle$ and $\psi^{(2)} \equiv \langle a_i^2 \rangle$. There are three distinct phases, namely of a Mott insulator ($\psi^{(1)}=\psi^{(2)}=0$), a conventional superfluid ($\psi^{(1)}\neq 0$, $\psi^{(2)}\neq 0$), and a pair superfluid ($\psi^{(1)}= 0$, $\psi^{(2)}\neq 0$). Markedly, there is an extended region of parameters in which the pair-superfluid phase is robust against single-photon hopping.

In Fig.~\ref{fig_supp:Gutzwiller_single_photon_hopping}c, we investigate in more details the pair-superfluid to conventional-superfluid transition. The single-particle hopping $J^{(1)}$ drives a second-order phase transition associated with the spontaneous breaking of the global $\mathbb{Z}_2$ symmetry where  $\psi^{(1)}$ continuously acquires  a finite value.
Beyond the order parameter, Fig.~\ref{fig_supp:Gutzwiller_single_photon_hopping}b illustrates how the Wigner function remains perfectly $\mathbb{Z}_2$ symmetric with interferences typical of cat states as long as one remains in the pair-superfluid phase, while asymmetric patterns appear once the system enters the $\mathbb{Z}_2$-broken phase.

\begin{figure*}[t]
	\centering
	\includegraphics[width=0.9\textwidth]{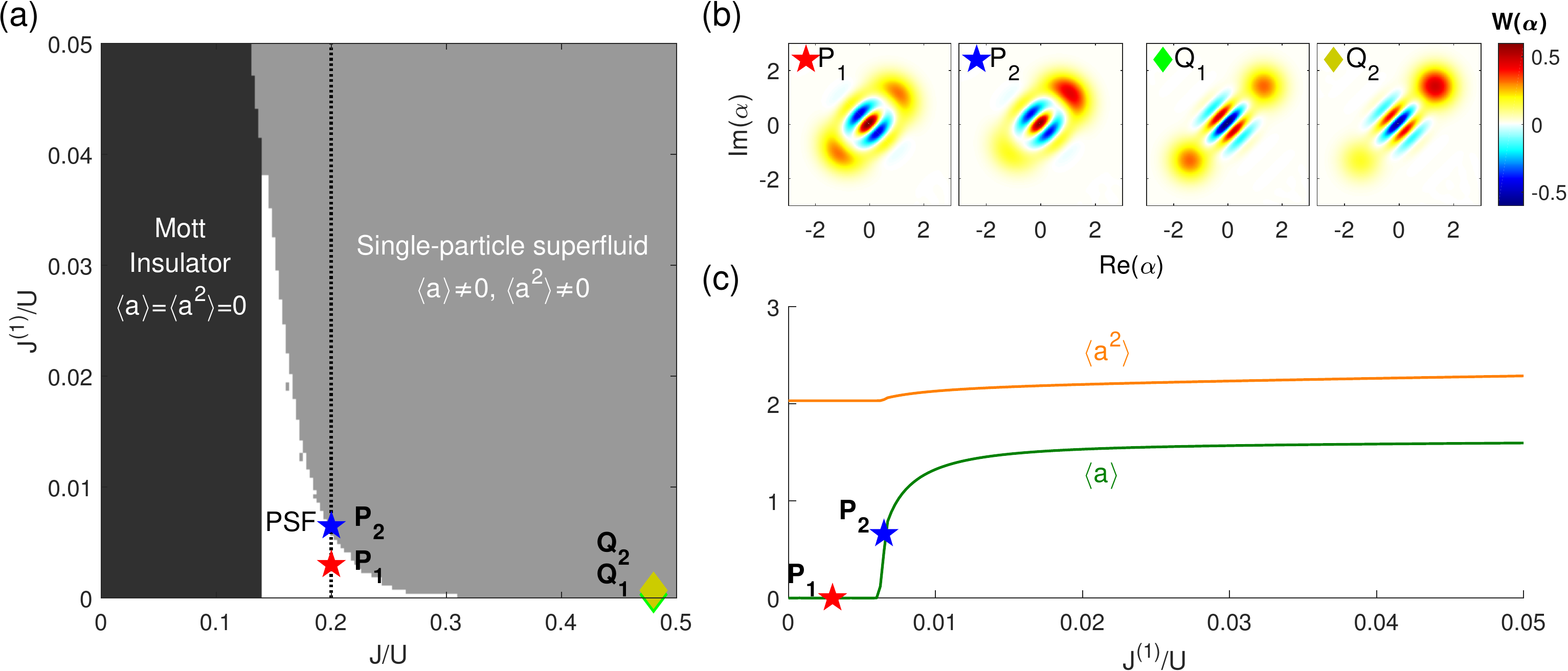} 
	\caption{Impact of single-particle hopping on the stability of pair superfluidity. \textbf{(a)}: Gutzwiller phase diagram as a function of the pair-hopping amplitude $J$ and the single-particle hopping amplitude $J^{(1)}$, both measured in units of $U$. \textbf{(b)}: Wigner functions $W(\alpha)$ of the Gutzwiller wave function $\ket{\psi_{\rm{GW}}}$ at the points ($P_1$), ($P_2$), ($Q_1$), ($Q_2$) indicated in panel (a), which are all located at $J^{(1)} > 0$ in order to assess the robustness of the $\mathbb{Z}_2$-symmetric phase. \textbf{(c)}: $\psi^{(1)} \equiv \langle a\rangle$ and $\psi^{(2)} \equiv \langle a^2\rangle$ as a function of  $J^{(1)}$ at fixed $J$ [indicated by a vertical dotted line in panel (a)]. Parameters: all simulations were performed at a fixed chemical potential $\mu/U=1.5$, except for the Wigner functions evaluated at  $Q_{1}$ and for $Q_{2}$ for which we chose  $\mu/U=0.15$ in order to reduce the gap $\Delta$ and to distinguish the PSF from the single-particle superfluid phase. For $P_{1}$ and $P_{2}$ one has $J/U=0.2$ and respectively $J^{(1)}/U=3\times 10^{-2},\,6.5\times 10^{-2}$, while for $Q_{1}$ and $Q_{2}$ one has $J/U=0.48$ and $J^{(1)}/U=5\times 10^{-5},\,8\times 10^{-5}$.
		\label{fig_supp:Gutzwiller_single_photon_hopping}}
\end{figure*}

\section*{\uppercase{Detection of many-body cat states}}
In this  section, we propose a detection scheme to probe the existence of the many-body cat states within the pair-superfluid phase. This approach is based on the measurement of a reduced Wigner quasi-probability distribution defined below in Equation~(\ref{eq_supp:red_wigner}), and thus is particularly suited for a photonic implementation of the Bose-Hubbard model. Although we focus here on the equilibrium scenario, the described detection scheme also applies to the driven-dissipative scenario. 

Let us first argue that one can not directly detect the many-body cat state structure at the single-site level if the ground-state  preparation scheme has preserved the $U(1)$ invariance of the model (\textit{e.g.}, if one has prepared the ground state via an adiabatic particle-number conserving scheme). In this case, the ground states $\ket{\psi_{N}^{P}}$ of $H_0$ are given by Eq.~(\ref{eq:ground-state_finite_mu})) in the main text and have a well-defined total particle number $N$. The single-site Wigner function of this state reads
\begin{equation}
W_i(\alpha)\equiv \langle\psi_N^P|\hat{W}_i(\alpha)|\psi_{N}^{P}\rangle
\end{equation}
with the operator $\hat{W}_i(\alpha)\equiv (2/\pi)\mathcal{D}_i(\alpha)\Pi_i\mathcal{D}_i(-\alpha)$,  where $\Pi_i$ and $\mathcal{D}_i(\alpha)$ are respectively the parity and displacement operators on site $i$, as defined in Ref.~\cite{Haroche}. The $U(1)$ invariance of $\ket{\psi_{N}^{P}}$ implies that $W_i(\alpha)$ is invariant under rotations in the complex plane: $W_i(\alpha \rme^{\textrm{i}\theta})=W_i(\alpha)$. Thus the anisotropic patterns related to the underlying cat-states cannot be detected. This difficulty can be lifted by computing the two-field Wigner function
\begin{equation}
W^{[2]}_{i,j}(\alpha,\beta)\equiv \langle\psi_N^P|\hat{W}_i(\alpha) \hat{W}_j(\beta)|\psi_{N}^{P}\rangle,
\end{equation}
where  $i\neq j$ are different lattices sites. An important long-range property of the wavefunction $\ket{\psi_{N}^{P}}$ is that $W^{[2]}_{i,j}(\alpha,\beta)$ only depends on the parities $P(i)$ and $P(j)$ and not on the positions of $i$ and $j$, nor their distance $|i-j|$. For example, if one chooses an even parity $P(i)=1$ throughout the lattice, then $W^{[2]}_{i,j}(\alpha,\beta)=W^{[2]}(\alpha,\beta)$ is fully independent of $i$ and $j$. 

As a consequence of the $U(1)$ invariance, the two-site Wigner function is invariant under global rotations:
$W^{[2]}_{i,j}(\alpha \rme^{\textrm{i}\theta},\beta \rme^{\textrm{i}\theta})=W^{[2]}_{i,j}(\alpha,\beta)$. However, $W^{[2]}(\alpha,\beta)$ may still depend on the relative phase between $\alpha$ and $\beta\in\mathbb{C}$. Thus, in order to characterize the existence of many-body cat states with locked relative phases, we suggest to measure the reduced Wigner function
\begin{equation}
\label{eq_supp:red_wigner}
W^{\rm{red}}_{i,j}(\alpha)=2\pi\int_0^{+\infty} \!\!\! \rm{d}\rho \, \rho \, W^{[2]}_{i,j}(\alpha,\rho),
\end{equation}
where the phase of the site $j$ has been fixed to a real value. Let us note that, as with a standard Wigner distribution, $W^{\rm{red}}_{i,j}(\alpha)$ is a real physical quantity verifying $\int d^2\alpha W^{\rm{red}}_{i,j}(\alpha)=1$. Moreover,  $W^{\rm{red}}_{i,j}(\alpha)=W^{\rm{red}}(\alpha)$ is independent of $i$ and $j$ when computed in the even parity sector on $\ket{\psi_N^P}$.

\begin{figure}[t]
	\centering
	\vspace{5mm}\includegraphics[width=0.6\columnwidth]{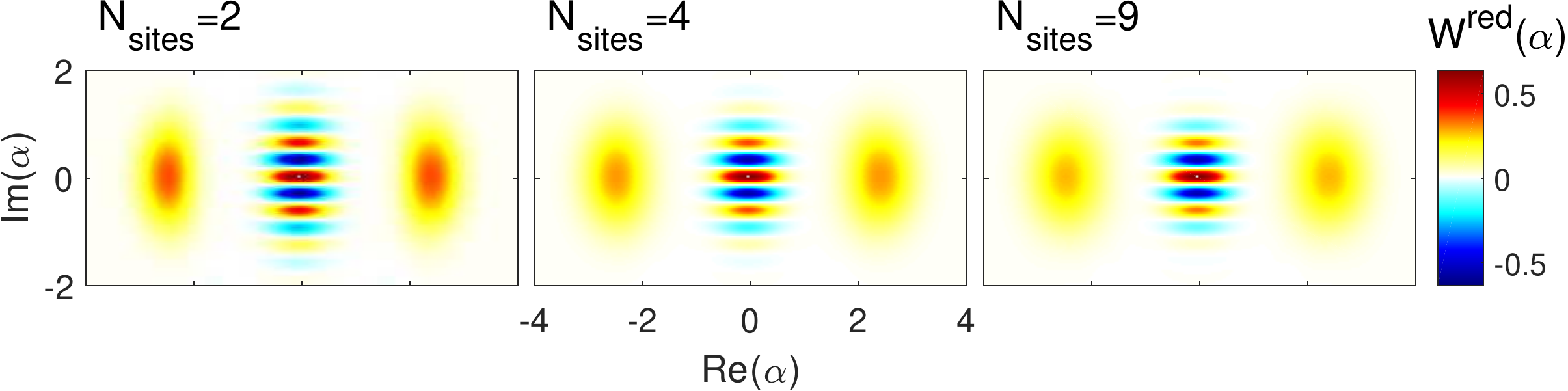} 
	\caption{\textbf{Detection of the many-body cat states.} Reduced Wigner function $W^{\rm{red}}_{i,j}(\alpha)$ of Eq.~(\ref{eq_supp:red_wigner}) computed on the N-particle ground-states $\ket{\psi_N^{P}}$ for various system sizes $N_{\rm{sites}}$. The computation was made at fixed density $n=N/N_{\rm{sites}}=6$ in the even parity sector $P(i)=1$, such that $W^{\rm{red}}_{i,j}(\alpha)=W^{\rm{red}}(\alpha)$ is independent of $i$ and $j$.
		\label{fig_supp:detection}}
\end{figure} 

The reduced Wigner function $W^{\rm{red}}_{i,j}(\alpha)$ is shown in Fig.~\ref{fig_supp:detection} for various system sizes at fixed density $n=N/N_{\rm{sites}}=6$. $W^{\rm{red}}_{i,j}(\alpha)$ possesses most of the essential characteristics of the Wigner function of a single cat state $|\mathcal{C}^{P(i)}(\alpha_0)\rangle$ with parity $P(i)$ and complex amplitude $\alpha_0\simeq\sqrt{n}$. In particular, the $\mathbb{Z}_2^{\rm{loc}}$ local invariance $\Pi_i\ket{\psi_{N}^{P}}=P(i)\ket{\psi_{N}^{P}}$ enforces  the reduced Wigner function to be invariant under $\pi$ rotations: $W^{\rm{red}}_{i,j}(-\alpha)=W^{\rm{red}}_{i,j}(\alpha)$. Moreover, the coherent nature of the resulting many-body phase is highlighted by the presence of interference fringes in the region close to $\alpha=0$, and one can show that $W^{\rm{red}}_{i,j}(0)=(2/\pi) P(i)$. However, there is a quenching in the density fluctuations for small system sizes, which progressively disappears when the number of sites is increased: this stems from the fact that rest of the lattice acts as a limited reservoir of particles for the $i$-th site. 

By means of exact diagonalization numerical methods, we checked that $W_{i,j}^{\rm{red}}(\alpha)$ does not significantly change for smaller hopping values $U/3\leq J\leq U/2$ and it already presents the structure of a cat state even though the N-particle ground-state does not completely coincide yet with $\ket{\psi_N^{P}}$. Moreover, we conjecture that the characterization procedure presented in this supplementary note extends to larger ensembles of sites, \textit{i.e.}, that the generalized reduced Wigner function 
\begin{equation}
W_{i_1,...,i_n}^{\rm{red}}(\alpha_1,...,\alpha_n)\equiv 2\pi\int\! \rm{d}\rho \, \rho \, W_{i_1,....,i_{n+1}}^{[n+1]}(\alpha_1,...,\alpha_n,\rho)
\end{equation}
is well approximated for a large $N_{\rm{sites}}$ by the product $\prod_{i=1}^{n}W_i^{\rm{cat}}(\alpha_i)$ of independent Wigner functions of cat states $\mathcal{C}^{P(i)}(\alpha_0)$ locked at the common amplitude $\alpha_0\simeq\sqrt{n}$. 

\section*{\uppercase{Transport-assisted dissipative stabilization of cat states arrays}}
A standard approach for the preparation of a single-cat state relies on the engineering of two-photon coherent drive and losses. In the perspective of large quantum registers, realizing such artificial reservoirs at each lattice site could turn to be prohibitively resource consuming. In this section, we discuss how a large array of cat states can be stabilized with only one single coherent drive, owing to the two-photon transport properties of the system. We consider the dynamics described by the following master equation:
\begin{equation}
\label{eq_supp:master_equation_transport}
\partial_t \rho=- \rmi \left[H_{\rm ph}+H_{\rm{d}},\rho \right]+\Gamma_{\rm{l}}\mathcal{D}[a_{0}^2](\rho),
\end{equation}
where $H_{\rm ph} = H_0 -\delta \sum_i a_i^\dagger a_i$. $H_0$ is the two-photon hopping Bose-Hubbard Hamiltonian introduced in Eq.~(\ref{eq:extended_BH})) of the main manuscript. The Hamiltonian $H_{\rm{d}}=\textrm{i}G(a_0^2-a_0^{\dagger 2})$ corresponds to a two-photon drive applied at a single site of the lattice, and $\delta=\omega_{\rm{d}}-\omega_{\rm{c}}$ is the detuning between the drive and cavity frequencies. $\Gamma_{\rm{l}}$ is the rate of an engineered two-photon loss applied at the same site.

Despite the obvious breaking of translational invariance by the driven-dissipative conditions, we show that transport properties of this model can contribute to creating an \emph{homogeneous} array of cat states for a certain choice of parameters. More precisely, we argue that for $J=U/2$ and $\delta=0$, the density matrices $\rho_{\infty}=\ket{\psi^{P}(\alpha_0)}\bra{\psi^{P}(\alpha_0)}$ are exact steady states of the model. Using previous notations, $\ket{\psi^{P}(\alpha)}=\bigotimes_i \ket{\mathcal{C}^{P(i)}(\alpha_0)}$ is a product of cat states with arbitrary local parities $P(i)$. While in the equilibrium case the amplitude of cat states was a free parameter, here $\alpha_0=\pm\sqrt{2G/\Gamma_{\rm{l}}}$ is now set by the drive-to-loss ratio. This is a straightforward consequence of previous results: for $J=U/2$ and zero detuning $\delta$, $H_{\rm{c}}\ket{\psi^{P}(\alpha)}=H_0\ket{\psi^{P}(\alpha)}=0$ and thus: $\left[H_{\rm ph},\rho_{\infty} \right]=0$. Moreover, as discussed in Ref.~\cite{Mirrahimi_cat}, the steady states of a single cavity subject to a two-photon coherent drive and to two-photon losses are cat states with arbitrary parities and a common amplitude set by $\alpha_0=\pm\sqrt{2G/\Gamma_{\rm{l}}}$. 


To conclude, despite the local character of the drive and dissipation, the transport properties in our system are efficient enough to restore translational invariance and generate an extended ensemble of cat states with identical amplitudes and free local parities at all sites.

\medskip

Interestingly enough, these results can also be generalized to the case of the circuit Hamiltonian in Eq.~(\ref{eq:BH_generalized}), where  pair interactions are explicitly mediated by auxiliary degrees of freedom. The dynamics are now described by the following master equation:
\begin{equation}
\partial_t \rho=- \rmi \left[\bar{H}+H_{\rm{d}},\rho \right]+\gamma_{\rm{l}}\mathcal{D}[b_{0,1}](\rho),
\end{equation}
where the Hamiltonian reads,  in the rotating frame $a_i\to a_i \textrm{exp}[\rmi(\omega_{\rm{c}}+\delta/2)t]$, $b_{i,j}\to b_{i,j} \textrm{exp}[\rmi(2\omega_{\rm{c}}+\delta)t]$,
\begin{equation}
\bar{H}=\sum_i \left\{-\delta a_{i}^{\dagger } a_{i}+\frac{U_0}{2}a_{i}^{\dagger 2} a_{i}^2\right\}+(-2\delta+\Delta)\sum_{\langle {i},{j}\rangle}b_{i,j}^\dagger b_{i,j}
-\sum_{\langle {i},{j}\rangle}\frac{\eta}{\sqrt{z}}\left[(a_{i}^{\dagger 2}+a_{j}^{\dagger 2}) b_{i,j}+(a_{i}^{2}+a_{j}^{2} ) b_{i,j}^\dagger\right]\,,
\end{equation}
and $\delta=\omega_d-2\omega_{\rm{c}}$ is the detuning between the drive  frequency and auxiliary resonator frequency.

One can show for the three-wave coupling value $\eta=\eta_c $ and  for a zero detuning $\delta=0$, that the density matrices $\rho_{\infty}=\ket{\psi^P(\alpha_0)}_{\rm{gen}}\bra{\psi^P(\alpha_0)}_{\rm{gen}}$ are exact steady states of the model. Here, the amplitude of the qubit cat states 
\begin{equation}
\alpha_0=\pm\left(\frac{\Delta}{2\sqrt{z}}\right)^{1/2}\sqrt{\frac{2F}{\gamma_{\rm{l}}}}
\end{equation}
and the coherent amplitude of the auxiliary resonators
\begin{equation}
\beta_0=\frac{2F}{\gamma_{\rm{l}}}=\frac{2\eta}{\Delta\sqrt{z}}\alpha_0^2
\end{equation}
are set by the single-photon drive-to-loss ratio of the auxiliary resonator, while the on-site parities $P(i)$ can vary arbitrarily. Therefore, an array of cat states living on the main lattice sites and of coherent states on the auxiliary resonators can be simply generated by applying single-photon drive and losses to a single lattice auxiliary site. 

\section*{\uppercase{Semiclassical analysis in the driven-dissipative case.}} We present here our semiclassical analysis of the driven-dissipative situation deep in the superfluid regime $\langle a_{i}^\dagger a_{i}\rangle\simeq |\langle a_{i}^2\rangle|\gg 1$. Any dissipative model expressed in terms of a master equation can be reformulated in terms in terms of Heisenberg equations of motions for the quantum field operators at the cost of including the external environment degrees of freedom in the Hamiltonian. In our case, this corresponds to including the two reservoirs responsible for the two-photon losses and for the dissipative pumping of the two-level systems.

For simplicity, let us first focus on the non-saturating regime $\Gamma_{\rm{p}}\gg n\Omega_{R}$ where the two-level systems constituting the emitters are perfectly inverted and respond linearly to the coupling to the two-photon field: under this assumption one can integrate out exactly the degrees of freedom of the various reservoirs  and of the two-level emitters, and derive a closed quantum Langevin for the photonic field (see Ref.~\cite{Lebreuilly_pseudo} for more details on this procedure). Assuming an homogeneous pair-superfluid order parameter $\psi(t)=\langle a_{i}^2\rangle (t)$ (we simplified the notation $\psi^{(2)}\to\psi$), and taking the average of the quantum Langevin equation one obtains the non-Markovian equation of evolution for $\psi(t)$: 
\begin{equation}
\label{eq:non-Markov_GP}
\frac{\textrm{d} \psi(t)}{\textrm{d} t} = -\rmi \left\lbrace 2\omega_{\rm{c}}+(4|\psi(t)|+1)\left[\frac{U}{2} -J\right]\right\rbrace \psi(t)+(4|\psi(t)|+1)\left[\int \rmd\tau \, \Gamma_{\rm{em}}(\tau)\psi(t-\tau)-
\frac{\Gamma_{\rm{l}}}{2}\psi(t)\right].
\end{equation}
$\Gamma_{\rm{em}}(\tau)=\theta(\tau)\int \frac{\rmd\omega}{2\pi}\mathcal{S}_{\rm{em}}(\omega) \rme^{-\rmi\omega\tau}$ is the memory kernel associated with the emission of photon pairs by the pumped two-level systems. One may search for a non-trivial steady-state solution of the form $\psi(t)=\psi_0 \rme^{-\rmi\omega_{\rm{PSF}}t}$ ($\psi_0\neq 0$), and we show that it satisfies the two relations
\begin{eqnarray}
\Gamma_{\rm{l}} &=& \mathcal{S}_{\rm{em}}(\omega_{\textrm{PSF}}),\label{eq:condensate_frequency}\\
\omega_{\textrm{PSF}} &=& 2\omega_{\rm{c}} +(4 |\psi_0|+1)\left\{\frac{U}{2}-J-\textrm{Im}\left[\Gamma_{\rm{em}}(\omega_{\textrm{PSF}})\right]\right\}.\nonumber 
\end{eqnarray}
The result in Eq.~(\ref{eq:SF-fraction_MF}) (main text) is obtained by neglecting the $+1$ term in the above equation (which is legitimate since $|\psi_0|\gg 1$), as well as the small Lamb shift $\textrm{Im}[\Gamma_{\rm{em}}(\omega_{\textrm{PSF}})]$ (since we worked in a  weakly-dissipative regime of parameters for which $\Gamma_{\rm{em}}^0\ll U,J$). This semiclassical model however does not include the tilting of the PSF domain described in the main manuscript, nor the presence of an upper bound for the PSF order parameter, which we show below to originate from saturation effects.

We now move to the full picture, including the two-level emitters and their saturation effect. In this case, one cannot directly integrate the Heisenberg equations of motion of the two-level emitters, and the evolution equation for $\psi(t)$ read
\begin{equation}
\label{eq:non-Markov_GP_sat}
\frac{\textrm{d} \psi(t)}{\textrm{d} t} = -\rmi \left\lbrace 2\omega_{\rm{c}}+(4|\psi(t)|+1)\left[\frac{U}{2} -J\right]\right\rbrace \psi(t)+(4|\psi(t)|+1)\left[-\rmi\Sigma(t)-
\frac{\Gamma_{\rm{l}}}{2}\psi(t)\right],
\end{equation}
including the coupling to the polarization $\Sigma(t)\equiv\langle \sigma^-_{i}\rangle (t)$ of the two-level emitters, which we assume to be spatially homogeneous. Eq.~(\ref{eq:non-Markov_GP_sat}) has to be completed by a dynamical model for the two-level emitters. We treat these as quantum degrees of freedom, leading to the following set of Bloch equations
\begin{eqnarray}
\label{eq:Bloch_equation}
\partial_t X(t)&=&-\Gamma_{\rm{p}}X(t)-2\rmi\Omega_{\rm{R}}[\psi(t)\Sigma^*(t)-\psi^*(t)\Sigma(t)]+\Gamma_{\rm{p}}\nonumber\\
\partial_t \Sigma(t)&=&\left[-\rmi\omega_{\rm{at}}-\frac{\Gamma_{\rm{p}}}{2}\right]\Sigma(t)+\rmi\Omega_{\rm{R}}\psi(t)X(t)\nonumber\\
\partial_t \Sigma^*(t)&=&\left[+\rmi\omega_{\rm{at}}-\frac{\Gamma_{\rm{p}}}{2}\right]\Sigma^*(t)-\rmi\Omega_{\rm{R}}\psi^*(t)X(t),
\end{eqnarray}
where $X(t)\equiv \langle \sigma^{z}\rangle(t)$ describes the population imbalance of the two-level emitters ($X=1$ corresponds to a perfect inversion of population), and the two-photon field $\psi(t)$ has been treated fully classically. The non-linearity makes the dynamics described by the Eqs.~(\ref{eq:non-Markov_GP_sat}) and~(\ref{eq:Bloch_equation}) relatively complicated. However,  one may by-pass the complex transient dynamics and obtain the following steady-state solution
\begin{eqnarray}
\psi(t)&=&\psi_0 \, \rme^{-\rmi\omega_{\rm{PSF}}t} \qquad (\psi_0\neq 0)\nonumber\\
\Sigma(t)&=& \Sigma_0 \, \rme^{-\rmi\omega_{\rm{PSF}}t} \nonumber\\
X(t)&=&X_0 \,,
\end{eqnarray}
which satisfies modified relations with respect to non-saturating case:
\begin{eqnarray}
\Gamma_{\rm{l}} &=& \frac{\mathcal{S}_{\rm{em}}(\omega_{\textrm{PSF}})}{1+s}\label{eq:condensate_frequency_sat}\,,\\
\omega_{\textrm{PSF}} &=& 2\omega_{\rm{c}} +(4 |\psi_0|+1)\left\{\frac{U}{2}-J-\frac{\textrm{Im}\left[\Gamma_{\rm{em}}(\omega_{\textrm{PSF}})\right]}{1+s}\right\}\,.\nonumber 
\end{eqnarray}
Here $s=2\frac{\mathcal{S}_{\rm{em}}(\omega_{\rm{PSF}})}{\Gamma_{\rm{p}}}|\psi_0|^2$ is the so-called saturation parameter, it quantifies how strongly the two-photon field affects the population inversion of the emitters as well as the photon pump power. A stability analysis of small perturbations around the numerical steady-state solutions of Eqs.~(\ref{eq:non-Markov_GP_sat}) and~(\ref{eq:Bloch_equation}) showed that at most one non-trivial solution ($\psi_0\neq 0$) was stable. This solution is the one represented in Fig.~\ref{fig:Gutzwiller_dissipative}c-d (main text). As before, upon neglecting the Lamb shift and the $+1$ term in the second equation, the nonlinear system~(\ref{eq:condensate_frequency_sat}) can be solved analytically for any choice of parameters. In particular, one obtains the estimates 
\begin{eqnarray}
\psi^{(2)}_{\rm{max}}&=&\sqrt{\frac{1}{2}\left(\frac{\Gamma_{\rm{p}}}{\Gamma_{\rm{l}}}-\frac{\Gamma_{\rm{p}}}{\Gamma_{\rm{em}}^0}\right)}\\
J_{c}&=&\frac{U}{2}-\frac{\delta}{2\psi^{(2)}_{\rm{max}}}
\end{eqnarray}
given in the main text for the upper bound $\psi_{\rm max}$ of the order parameter, and for its location $J_c$.


\end{document}